\documentclass[aps,prl,twocolumn,showpacs,floatfix,nobibnotes,superscriptaddress,longbibliography,nofootinbib]{revtex4-1}

\usepackage{amsmath}
\usepackage{inputenc}
\usepackage{amssymb}
\usepackage{graphicx}
\usepackage{verbatim}
\usepackage{epsfig}
\usepackage{bm}
\usepackage{color}
\usepackage{xcolor}
\usepackage{float}
\usepackage{dcolumn}
\usepackage{multirow} 
\usepackage{physics}
\usepackage{braket}

\usepackage[unicode=true,
  linktocpage,
  linkbordercolor={0.5 0.5 1},
  citebordercolor={0.5 1 0.5},
  colorlinks=true,
  linkcolor=blue,
  citecolor=blue,
  urlcolor=blue]{hyperref}

\usepackage{footnote}
\hypersetup{colorlinks=true,
    linkcolor=magenta,
    citecolor=blue,
    urlcolor=cyan,
    pdfpagemode=FullScreen,}

\usepackage{ulem}
\newcommand\redsout{\bgroup\markoverwith{\textcolor{blue}{\rule[0.5ex]{2pt}{1.4pt}}}\ULon}

\renewcommand{\emph}[1]{\textit{#1}}

\begin{document}


\title{$^{16}$O electroweak response functions from first principles}


\author{Bijaya Acharya}
\email{acharyabn@ornl.gov}
\affiliation{Physics Division, Oak Ridge National Laboratory,
Oak Ridge, TN 37831, USA} 

\author{Joanna E. Sobczyk}
\email{jsobczyk@uni-mainz.de}
\affiliation{Institut f\"ur Kernphysik and PRISMA$^+$ Cluster of Excellence, Johannes Gutenberg-Universit\"at, 55128 Mainz, Germany}

\author{Sonia Bacca}
\email{s.bacca@uni-mainz.de}
\affiliation{Institut f\"ur Kernphysik and PRISMA$^+$ Cluster of Excellence, Johannes Gutenberg-Universit\"at, 55128 Mainz, Germany}
\affiliation{Helmholtz-Institut Mainz, Johannes Gutenberg-Universit\"at Mainz, D-55099 Mainz, Germany}

\author{Gaute~Hagen}
\email{hageng@ornl.gov}
\affiliation{Physics Division, Oak Ridge National Laboratory,
Oak Ridge, TN 37831, USA} 
\affiliation{Department of Physics and Astronomy, University of Tennessee,
Knoxville, TN 37996, USA} 

\author{Weiguang Jiang}
\email{wjiang@uni-mainz.de}
\affiliation{Mainz Institut f\"ur Theoretical Physics and PRISMA$^+$ Cluster of Excellence, Johannes Gutenberg-Universit\"at, 55128 Mainz, Germany}

\begin{abstract}
We present calculations of various electroweak response functions for the $^{16}$O nucleus obtained using coupled-cluster theory in conjunction with the Lorentz integral transform method. We employ nuclear forces derived at next-to-leading order and next-to-next-to-leading order in chiral effective field theory and perform a Bayesian analysis to assess uncertainties. Our results are in good agreement with available electron-scattering data at $|{\bf q}| \approx 326$ MeV/c. Additionally, we provide several predictions for the weak response functions in the quasi-elastic peak region at $|{\bf q}| \approx 300$ and $400$ MeV/c, which are critical for long-baseline neutrino experiments.
\end{abstract}

\maketitle

\textit{Introduction---}
\label{introduction}
While the Standard Model accurately describes the interactions of leptons with elementary particles, the cross sections for leptons interacting with atomic nuclei involve significantly larger theoretical uncertainties. This is due to the fact that the nucleus is a complex many-body system composed of protons and neutrons held together by the strong force in a non-perturbative regime~\cite{fetter1971quantum, ring2004nuclear, mbte}. Recent advancements in nuclear theory have made substantial progress in calculating nuclear properties, including both static and, to a lesser extent, dynamic properties~\cite{Leidemann:2012hr, BaccaPastore}. The use of chiral effective field theory~\cite{Epelbaum:2008ga, machleidt2011, hammer2020} in combination with modern computational tools, often referred to as the {\it ab initio} approach~\cite{Hergert:2020bxy, Ekstrom:2022yea}, holds the greatest promise for quantifying and potentially reducing uncertainties.

In addition to providing a powerful means of probing nuclear structure, lepton-nucleus scattering is crucial for both astrophysics and particle physics. In astrophysics, the detection of neutrinos with energies on the order of 10 MeV originating from a supernova explosion in our Galaxy can provide insights into the progenitor and the explosion mechanism~\cite{DUNE:2020ypp}. In particle physics, the discovery that neutrinos have non-zero masses and undergo oscillations is among the most groundbreaking findings of this century. Understanding the oscillation mechanism through long-baseline neutrino experiments will be a major focus of physics research in the coming decades~\cite{DUNE,hyperk}. Unlike the $(e,e')$ case, where the initial and final electron energies are known, the incident neutrino energies must be reconstructed in these experiments. Accurate interpretation of the data on neutrino masses and mixing angles depends on reliable calculations of neutrino-nucleus scattering, particularly in the quasi-elastic kinematic regime at energies from a few hundred MeV to a few GeV~\cite{Ruso:2022qes}. Interestingly, in this range, precise electron-scattering experiments can be used to cross-check the implementation of the weak current operators, whose vector parts are the electromagnetic current operators that are probed by electron-scattering experiments. Furthermore, comparison with existing electron-scattering data also provides a validation of the treatment of the many-body dynamics.

Numerous $(e,e')$ cross-section measurements have been conducted in the past for stable nuclei across the nuclear chart~\cite{database}. Among the nuclear targets relevant for the long-baseline neutrino program, the most extensively studied nucleus is $^{12}$C~\cite{Benhar:1994hw,Ankowski:2020:PhysRevD.102.053001,Miha}, while $^{16}$O and $^{40}$Ar remain among the least explored. In the standard shell-model picture, both $^{12}$C and $^{16}$O are closed-shell nuclei, whereas $^{40}$Ar is an open-shell nucleus. $^{12}$C has been extensively investigated using Quantum Monte Carlo methods~\cite{Lovato_2016, Lovato:2017cux, Lovato_2020}. The $\alpha$-cluster structure of its ground and excited states~\cite{epelbaum2012,Otsuka:2022bcf} presents challenges for several other {\it ab initio} approaches. In contrast, inelastic scattering observables for the heavier nuclei $^{16}$O and $^{40}$Ar remain unexplored within an {\it ab initio} framework, although elastic scattering off $^{40}$Ar has been studied~\cite{Payne:2019wvy}.

In this Letter, we focus on the inelastic lepton-scattering observables of $^{16}$O for the first time within the {\it ab initio} approach. Our method of choice is the coupled-cluster theory~\cite{coester1958,coester1960,dean2004,bartlett2007,hagen2014}, a well-established many-body method which scales polynomially with the system size. We use nuclear forces derived from chiral effective field theory~\cite{Epelbaum:2008ga, machleidt2011, hammer2020}. Our approach links directly to the fundamental interactions of the Standard Model up to rigorous uncertainty estimates. This is an essential step toward a similar treatment of $^{40}$Ar, enabled by recent advances in {\it ab initio} methods for open-shell nuclei~\cite{miyagi2020,yao2020,Frosini:2021sxj,sun2024}.


\textit{Theory}---
The inclusive differential cross sections for the electron-scattering  $^{16}{\rm O}(e,e')$ reaction and the charged-current neutrino(anti-neutrino) scattering 
 $^{16}{\rm O}(\nu_e,e)$/$^{16}{\rm O}(\bar{\nu}_e,e^+)$ reactions 
can be written as~\cite{BaccaPastore,Acharya:2019fij}
 \begin{eqnarray}
 \label{eq:cross_section}
\nonumber
   \frac{d^2\sigma_e}{d\Omega d\epsilon^\prime} &=& \sigma_M \left[ \frac{Q^4}{q^4} R_{L}(\omega,q) +  \left(\tan^2 \frac{\theta}{2}+\frac{Q^2}{2q^2}\right)R_{T}(\omega,q)\right] \\
\nonumber 
    \frac{d^2\sigma_{\nu/\bar{\nu}}}{d\Omega  d\epsilon^\prime}&=& \frac{G_F^2}{2\pi^2}\,k^\prime E' \cos^2\frac{\theta}{2} \left [ R_{00}(\omega,q) + \frac{\omega^2}{q^2}R_{zz}(\omega,q)\right.\\  
    \nonumber
     &-& \frac{\omega}{q} R_{0z}(\omega,q) +\left(\tan^2 \frac{\theta}{2}+\frac{Q^2}{2q^2}\right) R_{xx} (\omega,q)\\
 &\mp& \left.
\tan\frac{\theta}{2}\sqrt{\tan^2\frac{\theta}{2}+
\frac{Q^2}{q^2}}
              R_{xy}(q,\omega) \right] \,.
\end{eqnarray}
Here, $\sigma_M$ is the Mott cross section, $G_F$ is the Fermi constant, $E'$ and $k'$ are the energy and momentum of the outgoing lepton, $+$/$-$ correspond to the neutrino and antineutrino, respectively, and $R_{\mu\nu}$ are the dynamical response functions that depend on the energy transfer $\omega$ and the momentum transfer ${\bf q}$, with $Q^2 = q^2 - \omega^2$. We take the momentum transfer $q = |{\bf q}|$ along the $z$ axis, such that $x$ and $y$ are the transverse components of the current operator.
The response functions are, in general, 
given by 
\begin{eqnarray}
\label{resp}
R_{\alpha\beta} (\omega,q) \!= \!\!\sum_{f} \langle \Psi_f |j_{\alpha} |\Psi_0 \rangle \! \langle \Psi_f |j_{\beta} |\Psi_0 \rangle^*
\delta\left(\omega+E_0-E_f\right)\!,
\end{eqnarray}
where $|\Psi_0 \rangle $ and $|\Psi_f\rangle$
are the nuclear ground and excited states with energy $E_0$ and $E_f$, respectively, and $j_{\alpha}$ is the four-vector current operator. For electron scattering, we separate the current in isoscalar $(S)$ and isovector $(V)$ parts as $j_{\alpha} = j^{(S)}_{\alpha} + j^{(V)}_{\alpha}$, and  we denote the only two non-vanishing response functions $R_{00}$ and $R_{xx}$ as $R_L$ and $R_T$, respectively.
The timelike and spacelike components of the isoscalar electromagnetic four-vector current $j^{(S)}_{\alpha}$ are given by 
\begin{equation}
\label{eq:isoscalarj0_v0}
 j^{(S)}_{0} = G_E^S(Q^2) e^{i\mathbf{q}\cdot\mathbf{r}_j}\frac{1}{2}\,,
\end{equation} 
and 
\begin{equation}
\label{eq:isoscalarjvec1}
 \mathbf{j}^{(S)} = \left( G_E^S(Q^2)\frac{\mathbf{\bar{p}}_j}{m} 
   - i \, G_M^S(Q^2)\frac{\mathbf{q}\times\bm{\sigma}_j}{2m} \right) e^{i\mathbf{q}\cdot\mathbf{r}_j}\frac{1}{2} \,,
\end{equation}
respectively. Here, $G_{E,M}^S\equiv G_{E,M}^p+G_{E,M}^n$ are the isoscalar electric and magnetic form factors, and $\mathbf{\bar{p}}_j=(\mathbf{p}_j^\prime+\mathbf{p}_j)/2 = \mathbf{p}_j+\mathbf{q}/2$ is the average of the initial and final momenta of the nucleon. 
In chiral effective field theory, Eq.~\eqref{eq:isoscalarj0_v0} contributes at leading order and Eq.~\eqref{eq:isoscalarjvec1} is suppressed by one power of the expansion parameter $p/\Lambda_\chi$, where $p$ is the typical momentum of the process and $\Lambda_\chi$ is the breakdown scale of the theory. The expressions for the isovector electromagnetic current operator, $j^{(V)}_{\alpha}$, 
can be obtained from the substitutions $
1/2 \rightarrow \tau_{j,z}/2$ and  $G_{E,M}^S \rightarrow G_{E,M}^V \equiv G_{E,M}^p-G_{E,M}^n\,$.

For neutrino scattering, the charge-changing weak current is taken as the sum of the vector term $j^{(\pm)}_{\alpha}$, which is related to $j^{(V)}_{\alpha}$ by a rotation in isospin space, $\tau_{j,z}/{2} \rightarrow \tau_{j,\pm} \equiv (\tau_{j,x} \pm i\,\tau_{j,y})/2\,,$
 and the axial term $j^{5(\pm)}_{\alpha}$. Here, the spacelike component 
\begin{equation}
\label{eq:1baxialcurrent1}
 \mathbf{j}^{5(\pm)}_{\alpha} = - G_A(Q^2) \, \left(\bm{\sigma}_j - \frac{\mathbf{q}\,\bm{\sigma}_j\cdot\mathbf{q}}{m_\pi^2+q^2}\right) \, e^{i\mathbf{q}\cdot\mathbf{r}_j} \, \frac{\tau_{j,\pm}}{2}
\end{equation} 
appears at the same order as Eq.~(\ref{eq:isoscalarj0_v0}), and the timelike (axial charge) component
\begin{equation}
\label{eq:1baxialcharge1}
 j_0^5 = - G_A(Q^2) \, \left(\bm{\sigma}_j \cdot \frac{\mathbf{\bar{p}}_j}{m} - \frac{\omega \,\bm{\sigma}_j\cdot\mathbf{q}}{m_\pi^2+q^2}\right) \, e^{i\mathbf{q}\cdot\mathbf{r}_j} \, \frac{\tau_{j,\pm}}{2} 
\end{equation}
is suppressed by a factor of $p/\Lambda_\chi$, i.e., it contributes at the same order as Eq.~\eqref{eq:isoscalarjvec1}. In Eqs.~\eqref{eq:1baxialcurrent1} and \eqref{eq:1baxialcharge1}, the pion-pole diagrams have been expressed in terms of the axial form factor $G_A(Q^2)$ by using a parametrization of the pseudoscalar form factor $G_P(Q^2)$ motivated by the Goldberger-Treiman relation \cite{Ericson:1988gk}. 
Two-body corrections  are not included in this work. In our power-counting (see Ref.~\cite{Acharya:2019fij}), they are suppressed by at least one chiral order compared to their one-body counterparts. 

 To solve for the groundstate of $^{16}$O, we first solve the Hartree-Fock equations starting from the intrinsic Hamiltonian with two- and three-nucleon forces from chiral effective field theory. In a second step we normal-order the Hamiltonian with respect to the Hartree-Fock state keeping up to two-body terms, i.e. we use the normal-ordered two-body approximation~\cite{hagen2007a,roth2012}. Finally, we include beyond-mean-field  correlations by employing the coupled-cluster (CC) method~\cite{hagen2014}. Notably, Eq.~(\ref{resp}) includes a sum over the excited states of the nucleus, whose explicit calculation for $^{16}$O is presently out of reach due to the many simultaneously-open channels in the continuum. We circumvent the problem by using the  Lorentz integral transform (LIT) method~\cite{efros2007}, which is based on integral-transforming  the response functions as 
\begin{equation}
   L_{\alpha\beta}(\sigma, \Gamma) = \frac{\Gamma}{\pi} \int d\omega \frac{R_{\alpha\beta}(\omega)}{(\omega - \sigma)^2 + \Gamma^2}=\langle \tilde{\Psi}_\alpha| \tilde{\Psi}_\beta\rangle,
   \label{lit}
\end{equation} 
where $\Gamma$ is the width and $\sigma$ is the centroid of the Lorentzian kernel. In this approach, $|\tilde{\Psi}_{\alpha,\beta}\rangle$ are bound-state-like objects and are, therefore, easier to calculate than $|\Psi_f\rangle$. 
From a numerical inversion of the transform we eventually retrieve the response functions, which include the full final state interaction (FSI)~\cite{efros2007}.
The LIT method has been extensively used in few-body systems, also for electron~\cite{Bacca:2006ji} and neutrino scattering~\cite{Gazit07}.
Here, we use it in conjunction with CC, adopting the so-called
LIT-CC method, which as been previously applied to photo-absorption and electron-nucleus reactions~\cite{bacca2013,bacca2014,miorelli2016, miorelli2018,simonis2019,Sobczyk:2021dwm,Sobczyk:2023sxh}.

\textit{Uncertainty quantification}---
We use the chiral interactions $\Delta$NLO$_{\rm GO}(450)$ and $\Delta$NNLO$_{\rm GO}(450)$ which include explicit $\Delta$-degrees of freedom at next-to leading and next-to-next-to leading order, respectively~\cite{jiang2020}. All results are calculated in the coupled-cluster singles and doubles approximation (CCSD)~\cite{bartlett2007,hagen2014} for an underlying harmonic oscillator frequency of $\hbar\Omega=14$ MeV and a model space of 15 major oscillator shells ($N_{\rm max}=2n+l=14$). A cut for matrix elements of three-body forces $e_{\rm 3max}=2(n_1+n_2+n_3)+l_1+l_2+l_3 \le16$ has also been imposed. We checked that our results are well converged with respect to the model-space size by varying $N_{\rm max}$ and $\hbar\Omega$.
For the inversion of the $L_{\alpha\beta}(\sigma, \Gamma)$, we follow the strategy already used in  Refs.~\cite{Sobczyk:2021dwm,Sobczyk:2023sxh}, where we expand the response function in terms of a linear combination of $N$ basis states that depend on a non-linear parameter. We determine the $N$ linear coefficients by a least-squares fit~\cite{efros2007}. We estimate the uncertainty associated with the inversion procedure by performing several LIT inversions: we use two different values of $\Gamma=$ 5, 10 MeV, vary $N$ from 6 to 8 and use $n_0=0.5$, $1.5$, a non-linear parameter of the basis states which governs the threshold behavior~\cite{Sobczyk:2021dwm}.  

\begin{figure}[h]
\includegraphics[width=1 \linewidth]{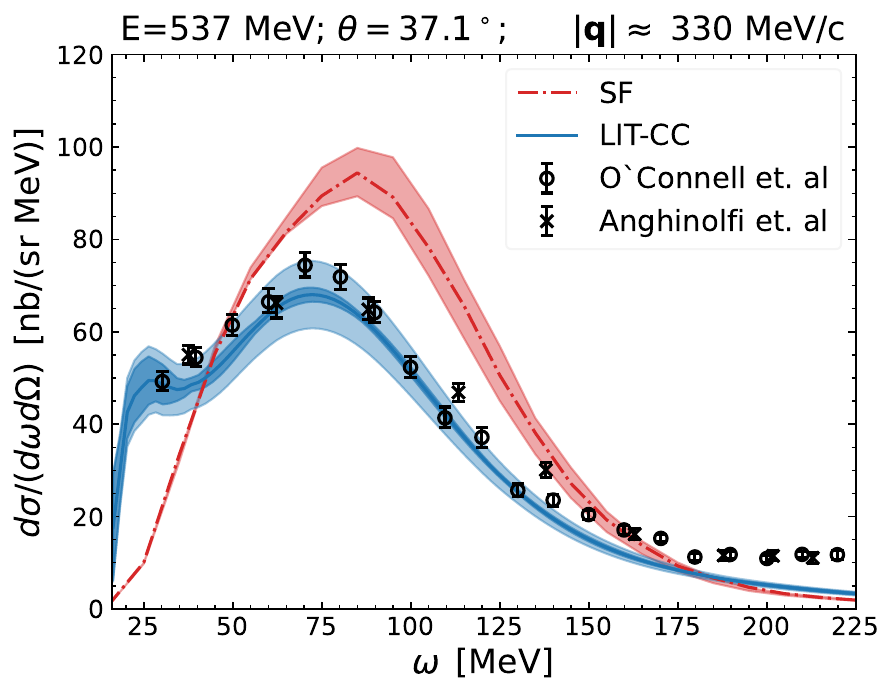}
\caption{$^{16}$O$(e,e')$ differential cross section as a function of the energy transfer: LIT-CC  and SF calculations with corresponding uncertainties (see text for details) compared to experimental data from Refs.~\cite{OConnell:1987ag,Anghinolfi:1996vm}. }
\label{FIG1}
\end{figure}

Following Ref.~\cite{Acharya:2021lrv}, we estimate the uncertainty of the chiral effective field theory truncation in powers of $p/\Lambda_\chi$, which is intrinsic to the nuclear potentials of our choice, using the Gaussian Process (GP) error model~\cite{Melendez:2019izc}. We compute observables also with the NNLO$_\mathrm{sat}$ potential~\cite{Ekstrom:2015rta} in the $\Delta$-less chiral theory and use it as a reference to ensure that the $\Delta$-full predictions can be expressed as smooth $\mathcal{O}(1)$ curves to which the GP model can be fit. This model allows us to validate our choices of $p$ and $\Lambda_\chi$ in addition to several other assumptions made in the uncertainty analysis (see Ref.~\cite{Acharya:2021lrv,Acharya:2024col}). Among the choices we explored, we found that $p\approx250$~MeV, $\Lambda_\chi\approx450$~MeV, yields the best GP fits under the diagnostic criteria discussed in Ref.~\cite{Melendez:2019izc}. We notice that the former is roughly equal to the Fermi momentum of a nucleon in the nucleus and the latter corresponds to the cutoff of the chiral interactions used in this work. 

 Overall, the uncertainty bands in our LIT-CC results represent a combination of the LIT inversion uncertainty and the chiral truncation uncertainty. The former, depicted as a darker band in the Figures, is dominant near the threshold, while the latter, combined in quadrature with the former to form the lighter band, becomes more significant at higher energies.  The LIT inversion uncertainty also includes the $\hbar\Omega$ dependence (estimated by varying it in the $14-20$ MeV range), which is negligibly small and therefore not shown separately.
 Effects from neglected triples excitations in the CC method and two-body currents are not included in the uncertainty analysis. In our power counting~\cite{Phillips:2016mov}, the latter are expected to be smaller than or comparable to the truncation error estimated from the potentials. 

\textit{Results}---
In Figure~\ref{FIG1}, we present our LIT-CC calculation of the $^{16}{\rm O}(e,e')$ reaction at $E=537$ MeV and $\theta=37.1^\circ$ (for momentum transfer $q\sim 330$ MeV/c), and compare it to calculations based on the spectral function (SF) formalism~\cite{Sobczyk:2023mey} and to data from Refs.~\cite{OConnell:1987ag,Anghinolfi:1996vm}.
 Experimentally,  the $^{12}{\rm C}$-$\alpha$ breakup threshold is located 7.162~MeV~\cite{Wang:2012eof}, slightly above the first $0^+$ excited state. The $0^+$ excited state is known to have a $\alpha$-cluster structure~\cite{epelbaum2014} which is not well described at the CCSD approximation level. Therefore, the $^{12}{\rm C}$-$\alpha$ is not the first break-up channel in our calculations, but rather the proton-emission channel. To determine this threshold energy, we compute the proton separation energy using the particle-removed CC method truncated at the two-hole-one-particle excitation level~\cite{gour2006,bartlett2007,hagen2014}, obtaining 10.6 MeV and 11.7 MeV for NNLO${\rm sat}$ and $\Delta$NNLO$_{\rm GO}(450)$, respectively,  in fair agreement with the experimental value of 12.1~MeV~\cite{Wang:2012eof}. We observe several bound excited states between 7 and 12 MeV with various multipolarities  other than $0^+$, which we remove  before inverting, similarly to what was done in Ref.~\cite{Sobczyk:2021dwm}.
Finally, to compare with the data taken at a fixed electron-scattering angle, we performed several calculations on a grid of momentum transfers, using a mesh spacing of 10 MeV, and interpolated linearly between the points.

In Figure~\ref{FIG1}, we see that the full treatment of the FSI achieved by the LIT-CC results gives an excellent description of data. While the SF approach, where the plane-wave impulse approximation is used, overestimate the data in the quasi-elastic peak, as expected in this momentum regime.  The uncertainty of the SF result comes from the propagation of the error in the spectral reconstruction of the SF computed within the coupled-cluster method. Typically, the SF approach works well for $q>500$ MeV/c~\cite{Sobczyk:2023mey}.

\begin{figure}[h]
\includegraphics[width=1 \linewidth]{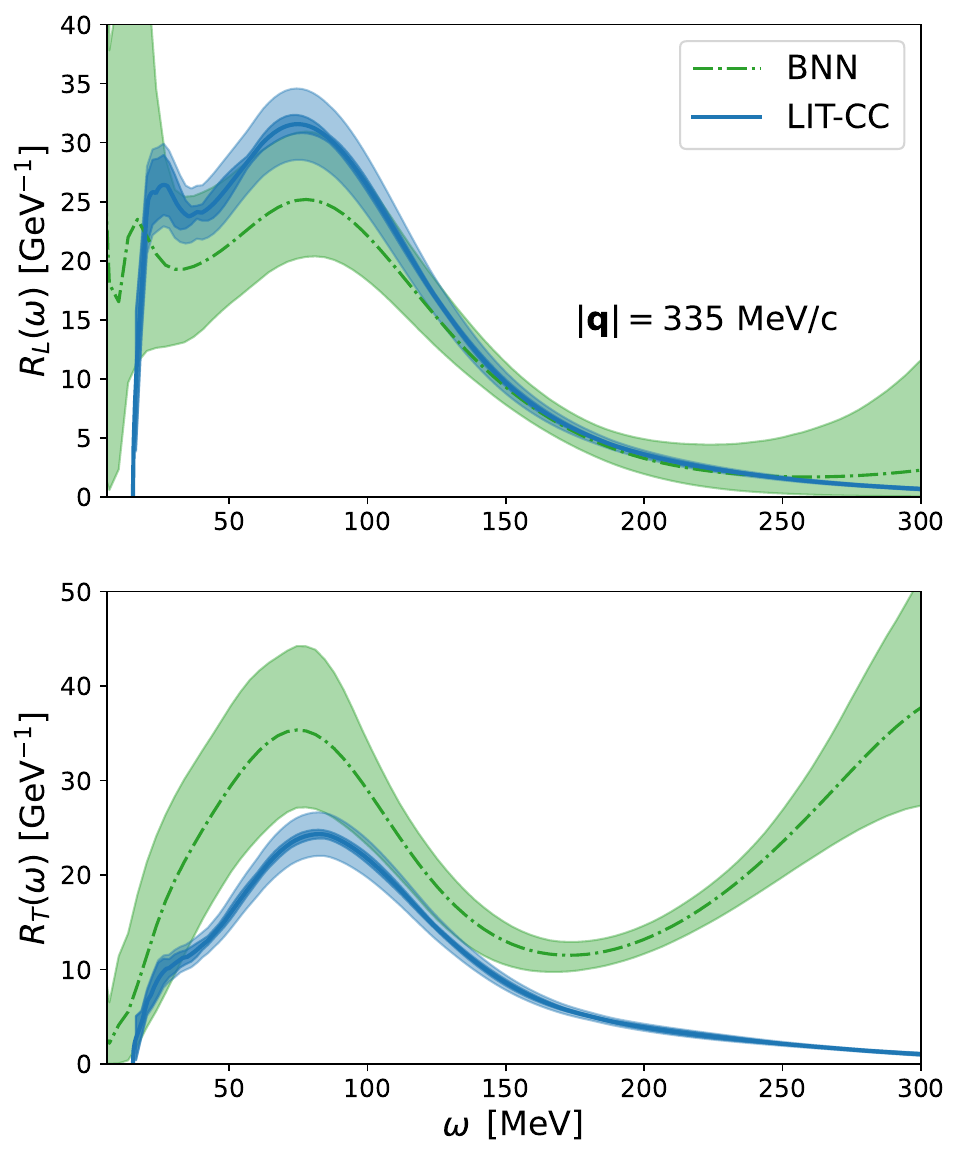}
\caption{$^{16}$O longitudinal (upper panel) and transverse (lower panel) response functions:  LIT-CC calculations with corresponding uncertainties (see text for details), compared with  the neural network predictions from Ref.~\cite{Sobczyk:ANN}.}
\label{FIG2}
\end{figure}

\begin{figure}[t]
\includegraphics[width=0.92 \linewidth]
{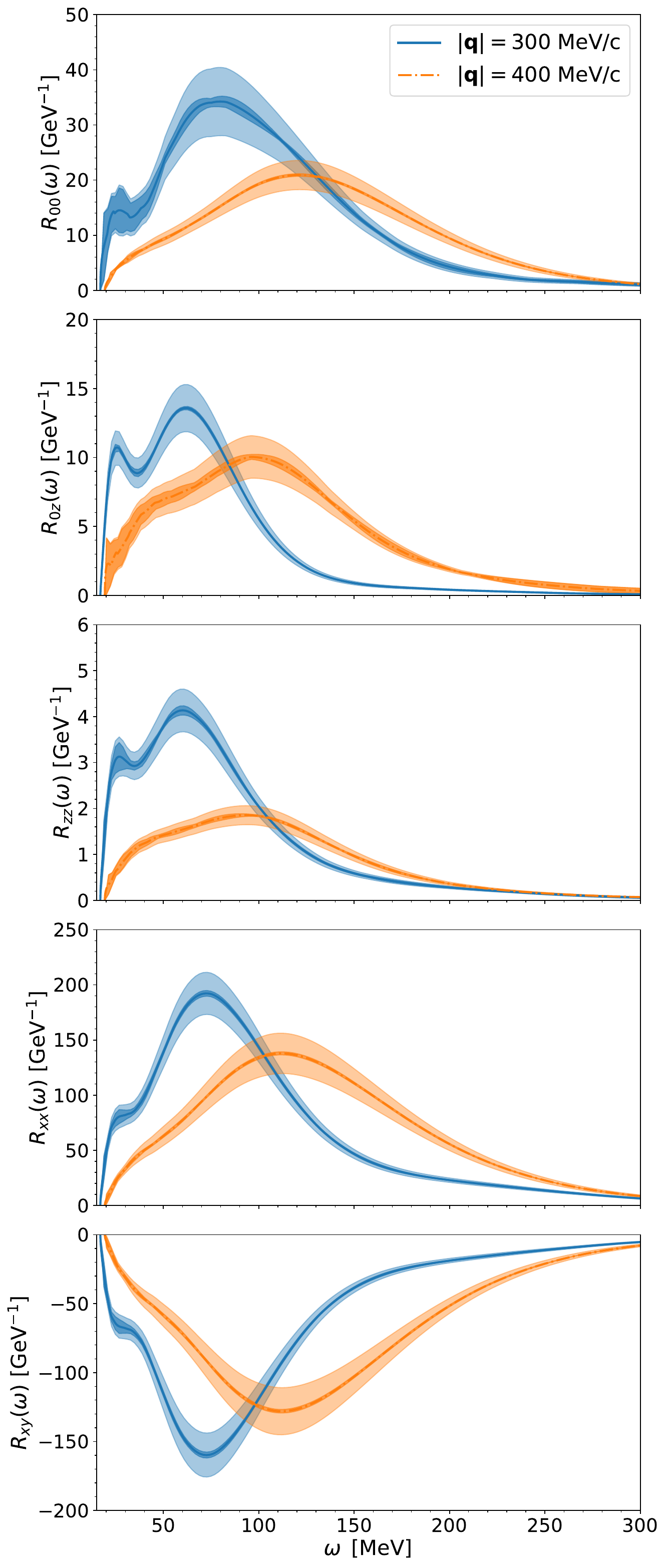}
\caption{The five different charge-current response functions related to the $^{16}$O$(\nu_l,l^-)$ reaction for $q=300,\, 400$ MeV/c with corresponding uncertainties (see text for details). }
\label{FIG3}
\end{figure}

The measurements in Refs.~\cite{OConnell:1987ag,Anghinolfi:1996vm} did not perform a Rosenbluth separation; therefore, we cannot individually compare the calculated electromagnetic response functions $R_L$ and $R_T$ with experimental data.
However, we can compare them with the prediction obtained from the Bayesian neutral network (BNN) devised in Ref.~\cite{Sobczyk:ANN} which is trained on the $(e,e')$ scattering data of several symmetric nuclei.
Such a comparison is presented in Fig.~\ref{FIG2} for the momentum transfer value $q=335$ MeV/c and energy transfer from 0 to 300 MeV. The BNN predictions are shown with the corresponding uncertainty, which is quite large in this case due to the scarsity of available data for $^{16}$O.

We observe that the BNN and LIT-CC results are compatible, particularly at low energy, while a discrepancy is seen for $R_T$ beyond the quasi-elastic peak, where the BNN prediction is consistently larger. This discrepancy can be understood in light of the fact that the LIT-CC calculation only includes one-body currents and does not account for pion production mechanisms. The neglected two-body currents would typically enhance the quasi-elastic peak~\cite{Lovato_2016}, as would the $\Delta$-isobar currents, which are expected to cause the rise in the response beyond $\omega = 200$ MeV predicted by the BNN. We note that the $\approx 15-20\%$ deficit in $R_T$ strength in the quasi-elastic peak region cannot be easily appreciated in Figure~\ref{FIG1}, as the cross-section is dominated by the longitudinal response at this kinematic setup.

We now turn to the discussion of the five $^{16}$O weak response functions: $R_{00}, R_{zz}, R_{0z}, R_{xx}$, and $R_{xy}$. These functions are computed for two momentum transfers, $q = 300$ MeV/c and $q = 400$ MeV/c, using 8 and 11 multipoles, respectively, to achieve convergence in the decomposition of the electroweak current operator.  When inverting the LIT, we impose the threshold energy of the $^{16}$O$(\nu_e,e)$p$^{15}$O channel
amounting to 12.5 MeV and 13.6 MeV for NNLO$_{\rm sat}$ and $\Delta$NNLO$_{\rm GO}(450)$, respectively, which compares relatively well with the experimental value of 14.4 MeV. As in the electromagnetic case, several excited states below the nucleon-emission threshold were identified and removed.

In Figure~\ref{FIG3}, similar to the electromagnetic case, the weak response functions exhibit a shoulder structure near the threshold. This feature was also observed in our LIT-CC calculations for $^4$He~\cite{Sobczyk:2021dwm} but not in the Quantum Monte Carlo calculations for $^{12}$C~\cite{Lovato_2020}. The difference may stem from variations in the employed Hamiltonian, the many-body method, or the nucleus itself. The position of the quasi-elastic peak in the weak response functions is approximately the same as in the electromagnetic case. As expected, and consistent with other findings~\cite{Lovato_2020}, the peak shifts to higher energies and its strength diminishes as the momentum transfer increases. Finally, the uncertainties in the results for $q = 300$ and $q = 400$ MeV/c are generally comparable in size, except for the $R_{00}$ response, where a larger uncertainty at $q = 300$ MeV/c is observed due to larger differences between the predictions of different chiral potentials.


\textit{Conclusions}--
Quasielastic lepton-nucleus scattering plays a crucial role in experimental programs aimed at studying fundamental neutrino properties, necessitating thorough calculations. Here, we compute and quantify the uncertainties associated with a set of properties related to $^{16}$O$(e,e')$ and $^{16}$O$(\nu_l,l^-)$ scattering in the {\it ab initio} LIT-CC framework using chiral forces. The observed agreement with available electron-scattering data and BNN-reconstructed responses provides an important validation of our approach. Additionally, computations of electroweak responses are a key input for future analyses of neutrino scattering on oxygen targets. For low-momentum transfers, where the plane-wave impulse approximation is not valid, these computationally intensive results could serve as a direct link to the analysis of experimental data if integrated into Monte Carlo event generators.

This work represents the first attempt to provide {\it ab initio} computations of electroweak reactions in the mass range of $A=16$ with quantified uncertainties. We estimate uncertainties from different sources: some originate from the many-body method and numerical procedures, which are dominant at low-energy transfer, while others stem from the truncation of the effective field theory used to model the strong force, which dominate at higher energies and are evaluated using modern statistical tools. Overall, our uncertainty estimates are reasonable and comparable to experimental errors when available. The potential to improve and reduce theoretical uncertainties makes our approach very promising. Future work in this area will involve studying higher-order many-body correlations and the effects of two-body currents.


\vspace{0.3cm}
\begin{acknowledgements}
We acknowledge useful discussions with Thomas Papenbrock.
This work was supported by the Deutsche
Forschungsgemeinschaft (DFG)
through the 
Cluster of Excellence ``Precision Physics, Fundamental
Interactions, and Structure of Matter" (PRISMA$^+$ EXC 2118/1, Project ID 390831469) and through the Collaborative Research Center ``Hadron and Nuclei as discovery tools" (Project ID 514321794), by the European Union’s Horizon 2020 research and innovation programme under the Marie Skłodowska-Curie grant agreement No.~101026014, by the Office of Nuclear Physics, U.S. Department of Energy under Contract No. DE-AC05-00OR22725 with Oak Ridge National Laboratory and under SciDAC-5 (NUCLEI collaboration), and by the Office of High Energy Physics, U.S. Department of Energy under Contract No. DE-AC02-07CH11359 through the Neutrino Theory Network Fellowship Program. Computer time was provided by the Innovative and Novel Computational Impact on Theory and Experiment (INCITE) program. This research used resources from the Oak Ridge Leadership Computing Facility located at ORNL, which is supported by the Office of Science of the Department of Energy under Contract No. DE-AC05-00OR22725, as well as the supercomputer MogonII at Johannes Gutenberg-Universit\"{a}t Mainz.

\end{acknowledgements}


\bibliography{biblio}

\begin{thebibliography}{60}%
\makeatletter
\providecommand \@ifxundefined [1]{%
 \@ifx{#1\undefined}
}%
\providecommand \@ifnum [1]{%
 \ifnum #1\expandafter \@firstoftwo
 \else \expandafter \@secondoftwo
 \fi
}%
\providecommand \@ifx [1]{%
 \ifx #1\expandafter \@firstoftwo
 \else \expandafter \@secondoftwo
 \fi
}%
\providecommand \natexlab [1]{#1}%
\providecommand \enquote  [1]{``#1''}%
\providecommand \bibnamefont  [1]{#1}%
\providecommand \bibfnamefont [1]{#1}%
\providecommand \citenamefont [1]{#1}%
\providecommand \href@noop [0]{\@secondoftwo}%
\providecommand \href [0]{\begingroup \@sanitize@url \@href}%
\providecommand \@href[1]{\@@startlink{#1}\@@href}%
\providecommand \@@href[1]{\endgroup#1\@@endlink}%
\providecommand \@sanitize@url [0]{\catcode `\\12\catcode `\$12\catcode
  `\&12\catcode `\#12\catcode `\^12\catcode `\_12\catcode `\%12\relax}%
\providecommand \@@startlink[1]{}%
\providecommand \@@endlink[0]{}%
\providecommand \url  [0]{\begingroup\@sanitize@url \@url }%
\providecommand \@url [1]{\endgroup\@href {#1}{\urlprefix }}%
\providecommand \urlprefix  [0]{URL }%
\providecommand \Eprint [0]{\href }%
\providecommand \doibase [0]{http://dx.doi.org/}%
\providecommand \selectlanguage [0]{\@gobble}%
\providecommand \bibinfo  [0]{\@secondoftwo}%
\providecommand \bibfield  [0]{\@secondoftwo}%
\providecommand \translation [1]{[#1]}%
\providecommand \BibitemOpen [0]{}%
\providecommand \bibitemStop [0]{}%
\providecommand \bibitemNoStop [0]{.\EOS\space}%
\providecommand \EOS [0]{\spacefactor3000\relax}%
\providecommand \BibitemShut  [1]{\csname bibitem#1\endcsname}%
\let\auto@bib@innerbib\@empty
\bibitem [{\citenamefont {Fetter}\ and\ \citenamefont
  {Walecka}(1971)}]{fetter1971quantum}%
  \BibitemOpen
  \bibfield  {author} {\bibinfo {author} {\bibfnamefont {Alexander~L.}\
  \bibnamefont {Fetter}}\ and\ \bibinfo {author} {\bibfnamefont {John~Dirk}\
  \bibnamefont {Walecka}},\ }\href@noop {} {\emph {\bibinfo {title} {Quantum
  Theory of Many-Particle Systems}}}\ (\bibinfo  {publisher} {McGraw-Hill},\
  \bibinfo {address} {New York},\ \bibinfo {year} {1971})\BibitemShut {NoStop}%
\bibitem [{\citenamefont {Ring}\ and\ \citenamefont
  {Schuck}(2004)}]{ring2004nuclear}%
  \BibitemOpen
  \bibfield  {author} {\bibinfo {author} {\bibfnamefont {Peter}\ \bibnamefont
  {Ring}}\ and\ \bibinfo {author} {\bibfnamefont {Peter}\ \bibnamefont
  {Schuck}},\ }\href {\doibase 10.1007/978-3-540-68077-9} {\emph {\bibinfo
  {title} {The Nuclear Many-Body Problem}}}\ (\bibinfo  {publisher}
  {Springer},\ \bibinfo {address} {Berlin, Heidelberg},\ \bibinfo {year}
  {2004})\BibitemShut {NoStop}%
\bibitem [{\citenamefont {Dickhoff}\ and\ \citenamefont
  {Van~Neck}(2005)}]{mbte}%
  \BibitemOpen
  \bibfield  {author} {\bibinfo {author} {\bibfnamefont {Willem~H}\
  \bibnamefont {Dickhoff}}\ and\ \bibinfo {author} {\bibfnamefont {Dimitri}\
  \bibnamefont {Van~Neck}},\ }\href {\doibase https://doi.org/10.1142/5804}
  {\emph {\bibinfo {title} {Many-Body Theory Exposed!}}}\ (\bibinfo
  {publisher} {World Scientific},\ \bibinfo {year} {2005})\BibitemShut
  {NoStop}%
\bibitem [{\citenamefont {Leidemann}\ and\ \citenamefont
  {Orlandini}(2013)}]{Leidemann:2012hr}%
  \BibitemOpen
  \bibfield  {author} {\bibinfo {author} {\bibfnamefont {Winfried}\
  \bibnamefont {Leidemann}}\ and\ \bibinfo {author} {\bibfnamefont
  {Giuseppina}\ \bibnamefont {Orlandini}},\ }\bibfield  {title} {\enquote
  {\bibinfo {title} {{Modern Ab Initio Approaches and Applications in
  Few-Nucleon Physics with A \ensuremath{>}= 4}},}\ }\href {\doibase
  10.1016/j.ppnp.2012.09.001} {\bibfield  {journal} {\bibinfo  {journal} {Prog.
  Part. Nucl. Phys.}\ }\textbf {\bibinfo {volume} {68}},\ \bibinfo {pages}
  {158--214} (\bibinfo {year} {2013})},\ \Eprint
  {http://arxiv.org/abs/1204.4617} {arXiv:1204.4617 [nucl-th]} \BibitemShut
  {NoStop}%
\bibitem [{\citenamefont {Bacca}\ and\ \citenamefont
  {Pastore}(2014)}]{BaccaPastore}%
  \BibitemOpen
  \bibfield  {author} {\bibinfo {author} {\bibfnamefont {Sonia}\ \bibnamefont
  {Bacca}}\ and\ \bibinfo {author} {\bibfnamefont {Saori}\ \bibnamefont
  {Pastore}},\ }\bibfield  {title} {\enquote {\bibinfo {title} {Electromagnetic
  reactions on light nuclei},}\ }\href {\doibase
  10.1088/0954-3899/41/12/123002} {\bibfield  {journal} {\bibinfo  {journal}
  {Journal of Physics G: Nuclear and Particle Physics}\ }\textbf {\bibinfo
  {volume} {41}},\ \bibinfo {pages} {123002} (\bibinfo {year}
  {2014})}\BibitemShut {NoStop}%
\bibitem [{\citenamefont {Epelbaum}\ \emph {et~al.}(2009)\citenamefont
  {Epelbaum}, \citenamefont {Hammer},\ and\ \citenamefont
  {Meissner}}]{Epelbaum:2008ga}%
  \BibitemOpen
  \bibfield  {author} {\bibinfo {author} {\bibfnamefont {Evgeny}\ \bibnamefont
  {Epelbaum}}, \bibinfo {author} {\bibfnamefont {Hans-Werner}\ \bibnamefont
  {Hammer}}, \ and\ \bibinfo {author} {\bibfnamefont {Ulf-G.}\ \bibnamefont
  {Meissner}},\ }\bibfield  {title} {\enquote {\bibinfo {title} {{Modern Theory
  of Nuclear Forces}},}\ }\href {\doibase 10.1103/RevModPhys.81.1773}
  {\bibfield  {journal} {\bibinfo  {journal} {Rev. Mod. Phys.}\ }\textbf
  {\bibinfo {volume} {81}},\ \bibinfo {pages} {1773--1825} (\bibinfo {year}
  {2009})},\ \Eprint {http://arxiv.org/abs/0811.1338} {arXiv:0811.1338
  [nucl-th]} \BibitemShut {NoStop}%
\bibitem [{\citenamefont {Machleidt}\ and\ \citenamefont
  {Entem}(2011)}]{machleidt2011}%
  \BibitemOpen
  \bibfield  {author} {\bibinfo {author} {\bibfnamefont {R.}~\bibnamefont
  {Machleidt}}\ and\ \bibinfo {author} {\bibfnamefont {D.R.}\ \bibnamefont
  {Entem}},\ }\bibfield  {title} {\enquote {\bibinfo {title} {Chiral effective
  field theory and nuclear forces},}\ }\href {\doibase
  https://doi.org/10.1016/j.physrep.2011.02.001} {\bibfield  {journal}
  {\bibinfo  {journal} {Physics Reports}\ }\textbf {\bibinfo {volume} {503}},\
  \bibinfo {pages} {1--75} (\bibinfo {year} {2011})}\BibitemShut {NoStop}%
\bibitem [{\citenamefont {Hammer}\ \emph {et~al.}(2020)\citenamefont {Hammer},
  \citenamefont {K\"onig},\ and\ \citenamefont {van Kolck}}]{hammer2020}%
  \BibitemOpen
  \bibfield  {author} {\bibinfo {author} {\bibfnamefont {H.-W.}\ \bibnamefont
  {Hammer}}, \bibinfo {author} {\bibfnamefont {Sebastian}\ \bibnamefont
  {K\"onig}}, \ and\ \bibinfo {author} {\bibfnamefont {U.}~\bibnamefont {van
  Kolck}},\ }\bibfield  {title} {\enquote {\bibinfo {title} {Nuclear effective
  field theory: Status and perspectives},}\ }\href {\doibase
  10.1103/RevModPhys.92.025004} {\bibfield  {journal} {\bibinfo  {journal}
  {Rev. Mod. Phys.}\ }\textbf {\bibinfo {volume} {92}},\ \bibinfo {pages}
  {025004} (\bibinfo {year} {2020})}\BibitemShut {NoStop}%
\bibitem [{\citenamefont {Hergert}(2020)}]{Hergert:2020bxy}%
  \BibitemOpen
  \bibfield  {author} {\bibinfo {author} {\bibfnamefont {H.}~\bibnamefont
  {Hergert}},\ }\bibfield  {title} {\enquote {\bibinfo {title} {{A Guided Tour
  of $ab$ $initio$ Nuclear Many-Body Theory}},}\ }\href {\doibase
  10.3389/fphy.2020.00379} {\bibfield  {journal} {\bibinfo  {journal} {Front.
  in Phys.}\ }\textbf {\bibinfo {volume} {8}},\ \bibinfo {pages} {379}
  (\bibinfo {year} {2020})},\ \Eprint {http://arxiv.org/abs/2008.05061}
  {arXiv:2008.05061 [nucl-th]} \BibitemShut {NoStop}%
\bibitem [{\citenamefont {Ekstr\"om}\ \emph {et~al.}(2023)\citenamefont
  {Ekstr\"om}, \citenamefont {Forss\'en}, \citenamefont {Hagen}, \citenamefont
  {Jansen}, \citenamefont {Jiang},\ and\ \citenamefont
  {Papenbrock}}]{Ekstrom:2022yea}%
  \BibitemOpen
  \bibfield  {author} {\bibinfo {author} {\bibfnamefont {A.}~\bibnamefont
  {Ekstr\"om}}, \bibinfo {author} {\bibfnamefont {C.}~\bibnamefont
  {Forss\'en}}, \bibinfo {author} {\bibfnamefont {G.}~\bibnamefont {Hagen}},
  \bibinfo {author} {\bibfnamefont {G.~R.}\ \bibnamefont {Jansen}}, \bibinfo
  {author} {\bibfnamefont {W.}~\bibnamefont {Jiang}}, \ and\ \bibinfo {author}
  {\bibfnamefont {T.}~\bibnamefont {Papenbrock}},\ }\bibfield  {title}
  {\enquote {\bibinfo {title} {{What is ab initio in nuclear theory?}}}\ }\href
  {\doibase 10.3389/fphy.2023.1129094} {\bibfield  {journal} {\bibinfo
  {journal} {Front. Phys.}\ }\textbf {\bibinfo {volume} {11}},\ \bibinfo
  {pages} {1129094} (\bibinfo {year} {2023})},\ \Eprint
  {http://arxiv.org/abs/2212.11064} {arXiv:2212.11064 [nucl-th]} \BibitemShut
  {NoStop}%
\bibitem [{\citenamefont {Abi}\ \emph {et~al.}(2020)\citenamefont {Abi} \emph
  {et~al.}}]{DUNE:2020ypp}%
  \BibitemOpen
  \bibfield  {author} {\bibinfo {author} {\bibfnamefont {Babak}\ \bibnamefont
  {Abi}} \emph {et~al.} (\bibinfo {collaboration} {DUNE}),\ }\bibfield  {title}
  {\enquote {\bibinfo {title} {{Deep Underground Neutrino Experiment (DUNE),
  Far Detector Technical Design Report, Volume II: DUNE Physics}},}\
  }\href@noop {} {\  (\bibinfo {year} {2020})},\ \Eprint
  {http://arxiv.org/abs/2002.03005} {arXiv:2002.03005 [hep-ex]} \BibitemShut
  {NoStop}%
\bibitem [{\citenamefont {Acciarri}\ \emph {et~al.}(2015)\citenamefont
  {Acciarri} \emph {et~al.}}]{DUNE}%
  \BibitemOpen
  \bibfield  {author} {\bibinfo {author} {\bibfnamefont {R.}~\bibnamefont
  {Acciarri}} \emph {et~al.} (\bibinfo {collaboration} {DUNE}),\ }\bibfield
  {title} {\enquote {\bibinfo {title} {{Long-Baseline Neutrino Facility (LBNF)
  and Deep Underground Neutrino Experiment (DUNE)}},}\ }\href@noop {} {\
  (\bibinfo {year} {2015})},\ \Eprint {http://arxiv.org/abs/1512.06148}
  {arXiv:1512.06148 [physics.ins-det]} \BibitemShut {NoStop}%
\bibitem [{\citenamefont {Abe}\ \emph {et~al.}(2015)\citenamefont {Abe} \emph
  {et~al.}}]{hyperk}%
  \BibitemOpen
  \bibfield  {author} {\bibinfo {author} {\bibfnamefont {K.}~\bibnamefont
  {Abe}} \emph {et~al.} (\bibinfo {collaboration} {Hyper-Kamiokande
  Proto-Collaboration}),\ }\bibfield  {title} {\enquote {\bibinfo {title}
  {{Physics potential of a long-baseline neutrino oscillation experiment using
  a J-PARC neutrino beam and Hyper-Kamiokande}},}\ }\href {\doibase
  10.1093/ptep/ptv061} {\bibfield  {journal} {\bibinfo  {journal} {PTEP}\
  }\textbf {\bibinfo {volume} {2015}},\ \bibinfo {pages} {053C02} (\bibinfo
  {year} {2015})}\BibitemShut {NoStop}%
\bibitem [{\citenamefont {Ruso}\ \emph {et~al.}(2022)\citenamefont {Ruso} \emph
  {et~al.}}]{Ruso:2022qes}%
  \BibitemOpen
  \bibfield  {author} {\bibinfo {author} {\bibfnamefont {L.~Alvarez}\
  \bibnamefont {Ruso}} \emph {et~al.},\ }\bibfield  {title} {\enquote {\bibinfo
  {title} {{Theoretical tools for neutrino scattering: interplay between
  lattice QCD, EFTs, nuclear physics, phenomenology, and neutrino event
  generators}},}\ }\href@noop {} {\  (\bibinfo {year} {2022})},\ \Eprint
  {http://arxiv.org/abs/2203.09030} {arXiv:2203.09030 [hep-ph]} \BibitemShut
  {NoStop}%
\bibitem [{dat()}]{database}%
  \BibitemOpen
  \href@noop {} {}\bibinfo {note}
  {Https://discovery.phys.virginia.edu/research/groups/qes-archive/}\BibitemShut
  {NoStop}%
\bibitem [{\citenamefont {Benhar}\ \emph {et~al.}(1994)\citenamefont {Benhar},
  \citenamefont {Fabrocini}, \citenamefont {Fantoni},\ and\ \citenamefont
  {Sick}}]{Benhar:1994hw}%
  \BibitemOpen
  \bibfield  {author} {\bibinfo {author} {\bibfnamefont {O.}~\bibnamefont
  {Benhar}}, \bibinfo {author} {\bibfnamefont {A.}~\bibnamefont {Fabrocini}},
  \bibinfo {author} {\bibfnamefont {S.}~\bibnamefont {Fantoni}}, \ and\
  \bibinfo {author} {\bibfnamefont {I.}~\bibnamefont {Sick}},\ }\bibfield
  {title} {\enquote {\bibinfo {title} {{Spectral function of finite nuclei and
  scattering of GeV electrons}},}\ }\href {\doibase
  10.1016/0375-9474(94)90920-2} {\bibfield  {journal} {\bibinfo  {journal}
  {Nucl. Phys. A}\ }\textbf {\bibinfo {volume} {579}},\ \bibinfo {pages}
  {493--517} (\bibinfo {year} {1994})}\BibitemShut {NoStop}%
\bibitem [{\citenamefont {Ankowski}\ and\ \citenamefont
  {Friedland}(2020)}]{Ankowski:2020:PhysRevD.102.053001}%
  \BibitemOpen
  \bibfield  {author} {\bibinfo {author} {\bibfnamefont {Artur~M.}\
  \bibnamefont {Ankowski}}\ and\ \bibinfo {author} {\bibfnamefont {Alexander}\
  \bibnamefont {Friedland}},\ }\bibfield  {title} {\enquote {\bibinfo {title}
  {Assessing the accuracy of the genie event generator with electron-scattering
  data},}\ }\href {\doibase 10.1103/PhysRevD.102.053001} {\bibfield  {journal}
  {\bibinfo  {journal} {Phys. Rev. D}\ }\textbf {\bibinfo {volume} {102}},\
  \bibinfo {pages} {053001} (\bibinfo {year} {2020})}\BibitemShut {NoStop}%
\bibitem [{\citenamefont {Mihovilovi\v{c}}\ \emph {et~al.}(2024)\citenamefont
  {Mihovilovi\v{c}} \emph {et~al.}}]{Miha}%
  \BibitemOpen
  \bibfield  {author} {\bibinfo {author} {\bibfnamefont {M.}~\bibnamefont
  {Mihovilovi\v{c}}} \emph {et~al.},\ }\bibfield  {title} {\enquote {\bibinfo
  {title} {{Measurement of the $\mathrm {{}^{12}C}(e,e')$ Cross Sections at
  $Q^2=0.8\,\textrm{GeV}^2/c^2$}},}\ }\href {\doibase
  10.1007/s00601-024-01944-y} {\bibfield  {journal} {\bibinfo  {journal} {Few
  Body Syst.}\ }\textbf {\bibinfo {volume} {65}},\ \bibinfo {pages} {78}
  (\bibinfo {year} {2024})},\ \Eprint {http://arxiv.org/abs/2406.16059}
  {arXiv:2406.16059 [nucl-ex]} \BibitemShut {NoStop}%
\bibitem [{\citenamefont {Lovato}\ \emph {et~al.}(2016)\citenamefont {Lovato},
  \citenamefont {Gandolfi}, \citenamefont {Carlson}, \citenamefont {Pieper},\
  and\ \citenamefont {Schiavilla}}]{Lovato_2016}%
  \BibitemOpen
  \bibfield  {author} {\bibinfo {author} {\bibfnamefont {A.}~\bibnamefont
  {Lovato}}, \bibinfo {author} {\bibfnamefont {S.}~\bibnamefont {Gandolfi}},
  \bibinfo {author} {\bibfnamefont {J.}~\bibnamefont {Carlson}}, \bibinfo
  {author} {\bibfnamefont {Steven~C.}\ \bibnamefont {Pieper}}, \ and\ \bibinfo
  {author} {\bibfnamefont {R.}~\bibnamefont {Schiavilla}},\ }\bibfield  {title}
  {\enquote {\bibinfo {title} {Electromagnetic response of $^{12}\mathrm{C}$: A
  first-principles calculation},}\ }\href {\doibase
  10.1103/PhysRevLett.117.082501} {\bibfield  {journal} {\bibinfo  {journal}
  {Phys. Rev. Lett.}\ }\textbf {\bibinfo {volume} {117}},\ \bibinfo {pages}
  {082501} (\bibinfo {year} {2016})}\BibitemShut {NoStop}%
\bibitem [{\citenamefont {Lovato}\ \emph {et~al.}(2018)\citenamefont {Lovato},
  \citenamefont {Gandolfi}, \citenamefont {Carlson}, \citenamefont {Lusk},
  \citenamefont {Pieper},\ and\ \citenamefont {Schiavilla}}]{Lovato:2017cux}%
  \BibitemOpen
  \bibfield  {author} {\bibinfo {author} {\bibfnamefont {A.}~\bibnamefont
  {Lovato}}, \bibinfo {author} {\bibfnamefont {S.}~\bibnamefont {Gandolfi}},
  \bibinfo {author} {\bibfnamefont {J.}~\bibnamefont {Carlson}}, \bibinfo
  {author} {\bibfnamefont {Ewing}\ \bibnamefont {Lusk}}, \bibinfo {author}
  {\bibfnamefont {Steven~C.}\ \bibnamefont {Pieper}}, \ and\ \bibinfo {author}
  {\bibfnamefont {R.}~\bibnamefont {Schiavilla}},\ }\bibfield  {title}
  {\enquote {\bibinfo {title} {{Quantum Monte Carlo calculation of
  neutral-current $\nu-^{12}C$ inclusive quasielastic scattering}},}\ }\href
  {\doibase 10.1103/PhysRevC.97.022502} {\bibfield  {journal} {\bibinfo
  {journal} {Phys. Rev. C}\ }\textbf {\bibinfo {volume} {97}},\ \bibinfo
  {pages} {022502} (\bibinfo {year} {2018})},\ \Eprint
  {http://arxiv.org/abs/1711.02047} {arXiv:1711.02047 [nucl-th]} \BibitemShut
  {NoStop}%
\bibitem [{\citenamefont {Lovato}\ \emph {et~al.}(2020)\citenamefont {Lovato},
  \citenamefont {Carlson}, \citenamefont {Gandolfi}, \citenamefont {Rocco},\
  and\ \citenamefont {Schiavilla}}]{Lovato_2020}%
  \BibitemOpen
  \bibfield  {author} {\bibinfo {author} {\bibfnamefont {A.}~\bibnamefont
  {Lovato}}, \bibinfo {author} {\bibfnamefont {J.}~\bibnamefont {Carlson}},
  \bibinfo {author} {\bibfnamefont {S.}~\bibnamefont {Gandolfi}}, \bibinfo
  {author} {\bibfnamefont {N.}~\bibnamefont {Rocco}}, \ and\ \bibinfo {author}
  {\bibfnamefont {R.}~\bibnamefont {Schiavilla}},\ }\bibfield  {title}
  {\enquote {\bibinfo {title} {Ab initio study of
  $({\ensuremath{\nu}}_{\ensuremath{\ell}},{\ensuremath{\ell}}^{\ensuremath{-}})$
  and
  $({\overline{\ensuremath{\nu}}}_{\ensuremath{\ell}},{\ensuremath{\ell}}^{+})$
  inclusive scattering in $^{12}\mathrm{C}$: Confronting the miniboone and t2k
  ccqe data},}\ }\href {\doibase 10.1103/PhysRevX.10.031068} {\bibfield
  {journal} {\bibinfo  {journal} {Phys. Rev. X}\ }\textbf {\bibinfo {volume}
  {10}},\ \bibinfo {pages} {031068} (\bibinfo {year} {2020})}\BibitemShut
  {NoStop}%
\bibitem [{\citenamefont {Epelbaum}\ \emph {et~al.}(2012)\citenamefont
  {Epelbaum}, \citenamefont {Krebs}, \citenamefont {L\"ahde}, \citenamefont
  {Lee},\ and\ \citenamefont {Mei\ss{}ner}}]{epelbaum2012}%
  \BibitemOpen
  \bibfield  {author} {\bibinfo {author} {\bibfnamefont {Evgeny}\ \bibnamefont
  {Epelbaum}}, \bibinfo {author} {\bibfnamefont {Hermann}\ \bibnamefont
  {Krebs}}, \bibinfo {author} {\bibfnamefont {Timo~A.}\ \bibnamefont
  {L\"ahde}}, \bibinfo {author} {\bibfnamefont {Dean}\ \bibnamefont {Lee}}, \
  and\ \bibinfo {author} {\bibfnamefont {Ulf-G.}\ \bibnamefont {Mei\ss{}ner}},\
  }\bibfield  {title} {\enquote {\bibinfo {title} {Structure and rotations of
  the hoyle state},}\ }\href {\doibase 10.1103/PhysRevLett.109.252501}
  {\bibfield  {journal} {\bibinfo  {journal} {Phys. Rev. Lett.}\ }\textbf
  {\bibinfo {volume} {109}},\ \bibinfo {pages} {252501} (\bibinfo {year}
  {2012})}\BibitemShut {NoStop}%
\bibitem [{\citenamefont {Otsuka}\ \emph {et~al.}(2022)\citenamefont {Otsuka},
  \citenamefont {Abe}, \citenamefont {Yoshida}, \citenamefont {Tsunoda},
  \citenamefont {Shimizu}, \citenamefont {Itagaki}, \citenamefont {Utsuno},
  \citenamefont {Vary}, \citenamefont {Maris},\ and\ \citenamefont
  {Ueno}}]{Otsuka:2022bcf}%
  \BibitemOpen
  \bibfield  {author} {\bibinfo {author} {\bibfnamefont {T.}~\bibnamefont
  {Otsuka}}, \bibinfo {author} {\bibfnamefont {T.}~\bibnamefont {Abe}},
  \bibinfo {author} {\bibfnamefont {T.}~\bibnamefont {Yoshida}}, \bibinfo
  {author} {\bibfnamefont {Y.}~\bibnamefont {Tsunoda}}, \bibinfo {author}
  {\bibfnamefont {N.}~\bibnamefont {Shimizu}}, \bibinfo {author} {\bibfnamefont
  {N.}~\bibnamefont {Itagaki}}, \bibinfo {author} {\bibfnamefont
  {Y.}~\bibnamefont {Utsuno}}, \bibinfo {author} {\bibfnamefont
  {J.}~\bibnamefont {Vary}}, \bibinfo {author} {\bibfnamefont {P.}~\bibnamefont
  {Maris}}, \ and\ \bibinfo {author} {\bibfnamefont {H.}~\bibnamefont {Ueno}},\
  }\bibfield  {title} {\enquote {\bibinfo {title}
  {{\ensuremath{\alpha}-Clustering in atomic nuclei from first principles with
  statistical learning and the Hoyle state character}},}\ }\href {\doibase
  10.1038/s41467-022-29582-0} {\bibfield  {journal} {\bibinfo  {journal}
  {Nature Commun.}\ }\textbf {\bibinfo {volume} {13}},\ \bibinfo {pages} {2234}
  (\bibinfo {year} {2022})}\BibitemShut {NoStop}%
\bibitem [{\citenamefont {Payne}\ \emph {et~al.}(2019)\citenamefont {Payne},
  \citenamefont {Bacca}, \citenamefont {Hagen}, \citenamefont {Jiang},\ and\
  \citenamefont {Papenbrock}}]{Payne:2019wvy}%
  \BibitemOpen
  \bibfield  {author} {\bibinfo {author} {\bibfnamefont {C.~G.}\ \bibnamefont
  {Payne}}, \bibinfo {author} {\bibfnamefont {S.}~\bibnamefont {Bacca}},
  \bibinfo {author} {\bibfnamefont {G.}~\bibnamefont {Hagen}}, \bibinfo
  {author} {\bibfnamefont {W.}~\bibnamefont {Jiang}}, \ and\ \bibinfo {author}
  {\bibfnamefont {T.}~\bibnamefont {Papenbrock}},\ }\bibfield  {title}
  {\enquote {\bibinfo {title} {{Coherent elastic neutrino-nucleus scattering on
  $^{40}$Ar from first principles}},}\ }\href {\doibase
  10.1103/PhysRevC.100.061304} {\bibfield  {journal} {\bibinfo  {journal}
  {Phys. Rev. C}\ }\textbf {\bibinfo {volume} {100}},\ \bibinfo {pages}
  {061304} (\bibinfo {year} {2019})},\ \Eprint
  {http://arxiv.org/abs/1908.09739} {arXiv:1908.09739 [nucl-th]} \BibitemShut
  {NoStop}%
\bibitem [{\citenamefont {Coester}(1958)}]{coester1958}%
  \BibitemOpen
  \bibfield  {author} {\bibinfo {author} {\bibfnamefont {F.}~\bibnamefont
  {Coester}},\ }\bibfield  {title} {\enquote {\bibinfo {title} {Bound states of
  a many-particle system},}\ }\href {\doibase 10.1016/0029-5582(58)90280-3}
  {\bibfield  {journal} {\bibinfo  {journal} {Nuclear Physics}\ }\textbf
  {\bibinfo {volume} {7}},\ \bibinfo {pages} {421 -- 424} (\bibinfo {year}
  {1958})}\BibitemShut {NoStop}%
\bibitem [{\citenamefont {Coester}\ and\ \citenamefont
  {K{\"u}mmel}(1960)}]{coester1960}%
  \BibitemOpen
  \bibfield  {author} {\bibinfo {author} {\bibfnamefont {F.}~\bibnamefont
  {Coester}}\ and\ \bibinfo {author} {\bibfnamefont {H.}~\bibnamefont
  {K{\"u}mmel}},\ }\bibfield  {title} {\enquote {\bibinfo {title} {Short-range
  correlations in nuclear wave functions},}\ }\href {\doibase
  10.1016/0029-5582(60)90140-1} {\bibfield  {journal} {\bibinfo  {journal}
  {Nuclear Physics}\ }\textbf {\bibinfo {volume} {17}},\ \bibinfo {pages} {477
  -- 485} (\bibinfo {year} {1960})}\BibitemShut {NoStop}%
\bibitem [{\citenamefont {Dean}\ and\ \citenamefont
  {Hjorth-Jensen}(2004)}]{dean2004}%
  \BibitemOpen
  \bibfield  {author} {\bibinfo {author} {\bibfnamefont {D.~J.}\ \bibnamefont
  {Dean}}\ and\ \bibinfo {author} {\bibfnamefont {M.}~\bibnamefont
  {Hjorth-Jensen}},\ }\bibfield  {title} {\enquote {\bibinfo {title}
  {Coupled-cluster approach to nuclear physics},}\ }\href {\doibase
  10.1103/PhysRevC.69.054320} {\bibfield  {journal} {\bibinfo  {journal} {Phys.
  Rev. C}\ }\textbf {\bibinfo {volume} {69}},\ \bibinfo {pages} {054320}
  (\bibinfo {year} {2004})}\BibitemShut {NoStop}%
\bibitem [{\citenamefont {Bartlett}\ and\ \citenamefont
  {Musia\l{}}(2007)}]{bartlett2007}%
  \BibitemOpen
  \bibfield  {author} {\bibinfo {author} {\bibfnamefont {Rodney~J.}\
  \bibnamefont {Bartlett}}\ and\ \bibinfo {author} {\bibfnamefont {Monika}\
  \bibnamefont {Musia\l{}}},\ }\bibfield  {title} {\enquote {\bibinfo {title}
  {Coupled-cluster theory in quantum chemistry},}\ }\href {\doibase
  10.1103/RevModPhys.79.291} {\bibfield  {journal} {\bibinfo  {journal} {Rev.
  Mod. Phys.}\ }\textbf {\bibinfo {volume} {79}},\ \bibinfo {pages} {291--352}
  (\bibinfo {year} {2007})}\BibitemShut {NoStop}%
\bibitem [{\citenamefont {Hagen}\ \emph {et~al.}(2014)\citenamefont {Hagen},
  \citenamefont {Papenbrock}, \citenamefont {Hjorth-Jensen},\ and\
  \citenamefont {Dean}}]{hagen2014}%
  \BibitemOpen
  \bibfield  {author} {\bibinfo {author} {\bibfnamefont {G.}~\bibnamefont
  {Hagen}}, \bibinfo {author} {\bibfnamefont {T.}~\bibnamefont {Papenbrock}},
  \bibinfo {author} {\bibfnamefont {M.}~\bibnamefont {Hjorth-Jensen}}, \ and\
  \bibinfo {author} {\bibfnamefont {D.~J.}\ \bibnamefont {Dean}},\ }\bibfield
  {title} {\enquote {\bibinfo {title} {Coupled-cluster computations of atomic
  nuclei},}\ }\href {\doibase 10.1088/0034-4885/77/9/096302} {\bibfield
  {journal} {\bibinfo  {journal} {Rep. Prog. Phys.}\ }\textbf {\bibinfo
  {volume} {77}},\ \bibinfo {pages} {096302} (\bibinfo {year}
  {2014})}\BibitemShut {NoStop}%
\bibitem [{\citenamefont {Miyagi}\ \emph {et~al.}(2020)\citenamefont {Miyagi},
  \citenamefont {Stroberg}, \citenamefont {Holt},\ and\ \citenamefont
  {Shimizu}}]{miyagi2020}%
  \BibitemOpen
  \bibfield  {author} {\bibinfo {author} {\bibfnamefont {T.}~\bibnamefont
  {Miyagi}}, \bibinfo {author} {\bibfnamefont {S.~R.}\ \bibnamefont
  {Stroberg}}, \bibinfo {author} {\bibfnamefont {J.~D.}\ \bibnamefont {Holt}},
  \ and\ \bibinfo {author} {\bibfnamefont {N.}~\bibnamefont {Shimizu}},\
  }\bibfield  {title} {\enquote {\bibinfo {title} {Ab initio multishell
  valence-space hamiltonians and the island of inversion},}\ }\href {\doibase
  10.1103/PhysRevC.102.034320} {\bibfield  {journal} {\bibinfo  {journal}
  {Phys. Rev. C}\ }\textbf {\bibinfo {volume} {102}},\ \bibinfo {pages}
  {034320} (\bibinfo {year} {2020})}\BibitemShut {NoStop}%
\bibitem [{\citenamefont {Yao}\ \emph {et~al.}(2020)\citenamefont {Yao},
  \citenamefont {Bally}, \citenamefont {Engel}, \citenamefont {Wirth},
  \citenamefont {Rodr\'{\i}guez},\ and\ \citenamefont {Hergert}}]{yao2020}%
  \BibitemOpen
  \bibfield  {author} {\bibinfo {author} {\bibfnamefont {J.~M.}\ \bibnamefont
  {Yao}}, \bibinfo {author} {\bibfnamefont {B.}~\bibnamefont {Bally}}, \bibinfo
  {author} {\bibfnamefont {J.}~\bibnamefont {Engel}}, \bibinfo {author}
  {\bibfnamefont {R.}~\bibnamefont {Wirth}}, \bibinfo {author} {\bibfnamefont
  {T.~R.}\ \bibnamefont {Rodr\'{\i}guez}}, \ and\ \bibinfo {author}
  {\bibfnamefont {H.}~\bibnamefont {Hergert}},\ }\bibfield  {title} {\enquote
  {\bibinfo {title} {Ab initio treatment of collective correlations and the
  neutrinoless double beta decay of $^{48}\mathrm{Ca}$},}\ }\href {\doibase
  10.1103/PhysRevLett.124.232501} {\bibfield  {journal} {\bibinfo  {journal}
  {Phys. Rev. Lett.}\ }\textbf {\bibinfo {volume} {124}},\ \bibinfo {pages}
  {232501} (\bibinfo {year} {2020})}\BibitemShut {NoStop}%
\bibitem [{\citenamefont {Frosini}\ \emph {et~al.}(2022)\citenamefont
  {Frosini}, \citenamefont {Duguet}, \citenamefont {Ebran}, \citenamefont
  {Bally}, \citenamefont {Mongelli}, \citenamefont {Rodr\'\i{}guez},
  \citenamefont {Roth},\ and\ \citenamefont {Som\`a}}]{Frosini:2021sxj}%
  \BibitemOpen
  \bibfield  {author} {\bibinfo {author} {\bibfnamefont {Mikael}\ \bibnamefont
  {Frosini}}, \bibinfo {author} {\bibfnamefont {Thomas}\ \bibnamefont
  {Duguet}}, \bibinfo {author} {\bibfnamefont {Jean-Paul}\ \bibnamefont
  {Ebran}}, \bibinfo {author} {\bibfnamefont {Benjamin}\ \bibnamefont {Bally}},
  \bibinfo {author} {\bibfnamefont {Tobias}\ \bibnamefont {Mongelli}}, \bibinfo
  {author} {\bibfnamefont {Tom\'as~R.}\ \bibnamefont {Rodr\'\i{}guez}},
  \bibinfo {author} {\bibfnamefont {Robert}\ \bibnamefont {Roth}}, \ and\
  \bibinfo {author} {\bibfnamefont {Vittorio}\ \bibnamefont {Som\`a}},\
  }\bibfield  {title} {\enquote {\bibinfo {title} {{Multi-reference many-body
  perturbation theory for nuclei: II. Ab initio study of neon isotopes via PGCM
  and IM-NCSM calculations}},}\ }\href {\doibase
  10.1140/epja/s10050-022-00693-y} {\bibfield  {journal} {\bibinfo  {journal}
  {Eur. Phys. J. A}\ }\textbf {\bibinfo {volume} {58}},\ \bibinfo {pages} {63}
  (\bibinfo {year} {2022})},\ \Eprint {http://arxiv.org/abs/2111.00797}
  {arXiv:2111.00797 [nucl-th]} \BibitemShut {NoStop}%
\bibitem [{\citenamefont {{Sun}}\ \emph {et~al.}(2024)\citenamefont {{Sun}},
  \citenamefont {{Ekstr{\"o}m}}, \citenamefont {{Forss{\'e}n}}, \citenamefont
  {{Hagen}}, \citenamefont {{Jansen}},\ and\ \citenamefont
  {{Papenbrock}}}]{sun2024}%
  \BibitemOpen
  \bibfield  {author} {\bibinfo {author} {\bibfnamefont {Z.~H.}\ \bibnamefont
  {{Sun}}}, \bibinfo {author} {\bibfnamefont {A.}~\bibnamefont
  {{Ekstr{\"o}m}}}, \bibinfo {author} {\bibfnamefont {C.}~\bibnamefont
  {{Forss{\'e}n}}}, \bibinfo {author} {\bibfnamefont {G.}~\bibnamefont
  {{Hagen}}}, \bibinfo {author} {\bibfnamefont {G.~R.}\ \bibnamefont
  {{Jansen}}}, \ and\ \bibinfo {author} {\bibfnamefont {T.}~\bibnamefont
  {{Papenbrock}}},\ }\bibfield  {title} {\enquote {\bibinfo {title}
  {{Multiscale physics of atomic nuclei from first principles}},}\ }\href
  {\doibase 10.48550/arXiv.2404.00058} {\bibfield  {journal} {\bibinfo
  {journal} {arXiv e-prints}\ ,\ \bibinfo {pages} {arXiv:2404.00058}} (\bibinfo
  {year} {2024})}\BibitemShut {NoStop}%
\bibitem [{\citenamefont {Acharya}\ and\ \citenamefont
  {Bacca}(2020)}]{Acharya:2019fij}%
  \BibitemOpen
  \bibfield  {author} {\bibinfo {author} {\bibfnamefont {Bijaya}\ \bibnamefont
  {Acharya}}\ and\ \bibinfo {author} {\bibfnamefont {Sonia}\ \bibnamefont
  {Bacca}},\ }\bibfield  {title} {\enquote {\bibinfo {title}
  {{Neutrino-deuteron scattering: Uncertainty quantification and new $L_{1,A}$
  constraints}},}\ }\href {\doibase 10.1103/PhysRevC.101.015505} {\bibfield
  {journal} {\bibinfo  {journal} {Phys. Rev. C}\ }\textbf {\bibinfo {volume}
  {101}},\ \bibinfo {pages} {015505} (\bibinfo {year} {2020})},\ \Eprint
  {http://arxiv.org/abs/1911.12659} {arXiv:1911.12659 [nucl-th]} \BibitemShut
  {NoStop}%
\bibitem [{\citenamefont {Ericson}\ and\ \citenamefont
  {Weise}(1988)}]{Ericson:1988gk}%
  \BibitemOpen
  \bibfield  {author} {\bibinfo {author} {\bibfnamefont {Torleif Erik~Oskar}\
  \bibnamefont {Ericson}}\ and\ \bibinfo {author} {\bibfnamefont
  {W.}~\bibnamefont {Weise}},\ }\href@noop {} {\emph {\bibinfo {title} {{Pions
  and Nuclei}}}}\ (\bibinfo  {publisher} {Clarendon Press},\ \bibinfo {address}
  {Oxford, UK},\ \bibinfo {year} {1988})\BibitemShut {NoStop}%
\bibitem [{\citenamefont {Hagen}\ \emph {et~al.}(2007)\citenamefont {Hagen},
  \citenamefont {Papenbrock}, \citenamefont {Dean}, \citenamefont {Schwenk},
  \citenamefont {Nogga}, \citenamefont {W\l{}och},\ and\ \citenamefont
  {Piecuch}}]{hagen2007a}%
  \BibitemOpen
  \bibfield  {author} {\bibinfo {author} {\bibfnamefont {G.}~\bibnamefont
  {Hagen}}, \bibinfo {author} {\bibfnamefont {T.}~\bibnamefont {Papenbrock}},
  \bibinfo {author} {\bibfnamefont {D.~J.}\ \bibnamefont {Dean}}, \bibinfo
  {author} {\bibfnamefont {A.}~\bibnamefont {Schwenk}}, \bibinfo {author}
  {\bibfnamefont {A.}~\bibnamefont {Nogga}}, \bibinfo {author} {\bibfnamefont
  {M.}~\bibnamefont {W\l{}och}}, \ and\ \bibinfo {author} {\bibfnamefont
  {P.}~\bibnamefont {Piecuch}},\ }\bibfield  {title} {\enquote {\bibinfo
  {title} {{Coupled-cluster theory for three-body Hamiltonians}},}\ }\href
  {\doibase 10.1103/PhysRevC.76.034302} {\bibfield  {journal} {\bibinfo
  {journal} {Phys. Rev. C}\ }\textbf {\bibinfo {volume} {76}},\ \bibinfo
  {pages} {034302} (\bibinfo {year} {2007})}\BibitemShut {NoStop}%
\bibitem [{\citenamefont {Roth}\ \emph {et~al.}(2012)\citenamefont {Roth},
  \citenamefont {Binder}, \citenamefont {Vobig}, \citenamefont {Calci},
  \citenamefont {Langhammer},\ and\ \citenamefont {Navr\'atil}}]{roth2012}%
  \BibitemOpen
  \bibfield  {author} {\bibinfo {author} {\bibfnamefont {Robert}\ \bibnamefont
  {Roth}}, \bibinfo {author} {\bibfnamefont {Sven}\ \bibnamefont {Binder}},
  \bibinfo {author} {\bibfnamefont {Klaus}\ \bibnamefont {Vobig}}, \bibinfo
  {author} {\bibfnamefont {Angelo}\ \bibnamefont {Calci}}, \bibinfo {author}
  {\bibfnamefont {Joachim}\ \bibnamefont {Langhammer}}, \ and\ \bibinfo
  {author} {\bibfnamefont {Petr}\ \bibnamefont {Navr\'atil}},\ }\bibfield
  {title} {\enquote {\bibinfo {title} {{Medium-Mass Nuclei with Normal-Ordered
  Chiral $NN\mathbf{+}3N$ Interactions}},}\ }\href {\doibase
  10.1103/PhysRevLett.109.052501} {\bibfield  {journal} {\bibinfo  {journal}
  {Phys. Rev. Lett.}\ }\textbf {\bibinfo {volume} {109}},\ \bibinfo {pages}
  {052501} (\bibinfo {year} {2012})}\BibitemShut {NoStop}%
\bibitem [{\citenamefont {Efros}\ \emph {et~al.}(2007)\citenamefont {Efros},
  \citenamefont {Leidemann}, \citenamefont {Orlandini},\ and\ \citenamefont
  {Barnea}}]{efros2007}%
  \BibitemOpen
  \bibfield  {author} {\bibinfo {author} {\bibfnamefont {V~D}\ \bibnamefont
  {Efros}}, \bibinfo {author} {\bibfnamefont {W}~\bibnamefont {Leidemann}},
  \bibinfo {author} {\bibfnamefont {G}~\bibnamefont {Orlandini}}, \ and\
  \bibinfo {author} {\bibfnamefont {N}~\bibnamefont {Barnea}},\ }\bibfield
  {title} {\enquote {\bibinfo {title} {The lorentz integral transform (lit)
  method and its applications to perturbation-induced reactions},}\ }\href
  {http://stacks.iop.org/0954-3899/34/i=12/a=R02} {\bibfield  {journal}
  {\bibinfo  {journal} {Journal of Physics G: Nuclear and Particle Physics}\
  }\textbf {\bibinfo {volume} {34}},\ \bibinfo {pages} {R459} (\bibinfo {year}
  {2007})}\BibitemShut {NoStop}%
\bibitem [{\citenamefont {Bacca}\ \emph {et~al.}(2007)\citenamefont {Bacca},
  \citenamefont {Arenhoevel}, \citenamefont {Barnea}, \citenamefont
  {Leidemann},\ and\ \citenamefont {Orlandini}}]{Bacca:2006ji}%
  \BibitemOpen
  \bibfield  {author} {\bibinfo {author} {\bibfnamefont {S.}~\bibnamefont
  {Bacca}}, \bibinfo {author} {\bibfnamefont {H.}~\bibnamefont {Arenhoevel}},
  \bibinfo {author} {\bibfnamefont {N.}~\bibnamefont {Barnea}}, \bibinfo
  {author} {\bibfnamefont {W.}~\bibnamefont {Leidemann}}, \ and\ \bibinfo
  {author} {\bibfnamefont {G.}~\bibnamefont {Orlandini}},\ }\bibfield  {title}
  {\enquote {\bibinfo {title} {{Inclusive electron scattering off He-4}},}\
  }\href {\doibase 10.1016/j.nuclphysa.2007.03.065} {\bibfield  {journal}
  {\bibinfo  {journal} {Phys. Rev. C}\ }\textbf {\bibinfo {volume} {76}},\
  \bibinfo {pages} {014003} (\bibinfo {year} {2007})},\ \Eprint
  {http://arxiv.org/abs/nucl-th/0612009} {arXiv:nucl-th/0612009} \BibitemShut
  {NoStop}%
\bibitem [{\citenamefont {Gazit}\ and\ \citenamefont {Barnea}(2007)}]{Gazit07}%
  \BibitemOpen
  \bibfield  {author} {\bibinfo {author} {\bibfnamefont {Doron}\ \bibnamefont
  {Gazit}}\ and\ \bibinfo {author} {\bibfnamefont {Nir}\ \bibnamefont
  {Barnea}},\ }\bibfield  {title} {\enquote {\bibinfo {title} {Low-energy
  inelastic neutrino reactions on $^{4}\mathrm{He}$},}\ }\href {\doibase
  10.1103/PhysRevLett.98.192501} {\bibfield  {journal} {\bibinfo  {journal}
  {Phys. Rev. Lett.}\ }\textbf {\bibinfo {volume} {98}},\ \bibinfo {pages}
  {192501} (\bibinfo {year} {2007})}\BibitemShut {NoStop}%
\bibitem [{\citenamefont {Bacca}\ \emph {et~al.}(2013)\citenamefont {Bacca},
  \citenamefont {Barnea}, \citenamefont {Hagen}, \citenamefont {Orlandini},\
  and\ \citenamefont {Papenbrock}}]{bacca2013}%
  \BibitemOpen
  \bibfield  {author} {\bibinfo {author} {\bibfnamefont {S.}~\bibnamefont
  {Bacca}}, \bibinfo {author} {\bibfnamefont {N.}~\bibnamefont {Barnea}},
  \bibinfo {author} {\bibfnamefont {G.}~\bibnamefont {Hagen}}, \bibinfo
  {author} {\bibfnamefont {G.}~\bibnamefont {Orlandini}}, \ and\ \bibinfo
  {author} {\bibfnamefont {T.}~\bibnamefont {Papenbrock}},\ }\bibfield  {title}
  {\enquote {\bibinfo {title} {First principles description of the giant dipole
  resonance in $^{16}\mathbf{O}$},}\ }\href {\doibase
  10.1103/PhysRevLett.111.122502} {\bibfield  {journal} {\bibinfo  {journal}
  {Phys. Rev. Lett.}\ }\textbf {\bibinfo {volume} {111}},\ \bibinfo {pages}
  {122502} (\bibinfo {year} {2013})}\BibitemShut {NoStop}%
\bibitem [{\citenamefont {{Bacca}}\ \emph {et~al.}(2014)\citenamefont
  {{Bacca}}, \citenamefont {{Barnea}}, \citenamefont {{Hagen}}, \citenamefont
  {{Miorelli}}, \citenamefont {{Orlandini}},\ and\ \citenamefont
  {{Papenbrock}}}]{bacca2014}%
  \BibitemOpen
  \bibfield  {author} {\bibinfo {author} {\bibfnamefont {S.}~\bibnamefont
  {{Bacca}}}, \bibinfo {author} {\bibfnamefont {N.}~\bibnamefont {{Barnea}}},
  \bibinfo {author} {\bibfnamefont {G.}~\bibnamefont {{Hagen}}}, \bibinfo
  {author} {\bibfnamefont {M.}~\bibnamefont {{Miorelli}}}, \bibinfo {author}
  {\bibfnamefont {G.}~\bibnamefont {{Orlandini}}}, \ and\ \bibinfo {author}
  {\bibfnamefont {T.}~\bibnamefont {{Papenbrock}}},\ }\bibfield  {title}
  {\enquote {\bibinfo {title} {{Giant and pigmy dipole resonances in 4He,
  16,22O, and 40Ca from chiral nucleon-nucleon interactions}},}\ }\href
  {http://adsabs.harvard.edu/abs/2014arXiv1410.2258B} {\bibfield  {journal}
  {\bibinfo  {journal} {ArXiv e-prints}\ } (\bibinfo {year} {2014})},\ \Eprint
  {http://arxiv.org/abs/1410.2258} {arXiv:1410.2258 [nucl-th]} \BibitemShut
  {NoStop}%
\bibitem [{\citenamefont {Miorelli}\ \emph {et~al.}(2016)\citenamefont
  {Miorelli}, \citenamefont {Bacca}, \citenamefont {Barnea}, \citenamefont
  {Hagen}, \citenamefont {Jansen}, \citenamefont {Orlandini},\ and\
  \citenamefont {Papenbrock}}]{miorelli2016}%
  \BibitemOpen
  \bibfield  {author} {\bibinfo {author} {\bibfnamefont {M.}~\bibnamefont
  {Miorelli}}, \bibinfo {author} {\bibfnamefont {S.}~\bibnamefont {Bacca}},
  \bibinfo {author} {\bibfnamefont {N.}~\bibnamefont {Barnea}}, \bibinfo
  {author} {\bibfnamefont {G.}~\bibnamefont {Hagen}}, \bibinfo {author}
  {\bibfnamefont {G.~R.}\ \bibnamefont {Jansen}}, \bibinfo {author}
  {\bibfnamefont {G.}~\bibnamefont {Orlandini}}, \ and\ \bibinfo {author}
  {\bibfnamefont {T.}~\bibnamefont {Papenbrock}},\ }\bibfield  {title}
  {\enquote {\bibinfo {title} {Electric dipole polarizability from first
  principles calculations},}\ }\href {\doibase 10.1103/PhysRevC.94.034317}
  {\bibfield  {journal} {\bibinfo  {journal} {Phys. Rev. C}\ }\textbf {\bibinfo
  {volume} {94}},\ \bibinfo {pages} {034317} (\bibinfo {year}
  {2016})}\BibitemShut {NoStop}%
\bibitem [{\citenamefont {Miorelli}\ \emph {et~al.}(2018)\citenamefont
  {Miorelli}, \citenamefont {Bacca}, \citenamefont {Hagen},\ and\ \citenamefont
  {Papenbrock}}]{miorelli2018}%
  \BibitemOpen
  \bibfield  {author} {\bibinfo {author} {\bibfnamefont {M.}~\bibnamefont
  {Miorelli}}, \bibinfo {author} {\bibfnamefont {S.}~\bibnamefont {Bacca}},
  \bibinfo {author} {\bibfnamefont {G.}~\bibnamefont {Hagen}}, \ and\ \bibinfo
  {author} {\bibfnamefont {T.}~\bibnamefont {Papenbrock}},\ }\bibfield  {title}
  {\enquote {\bibinfo {title} {Computing the dipole polarizability of
  $^{48}\mathrm{Ca}$ with increased precision},}\ }\href {\doibase
  10.1103/PhysRevC.98.014324} {\bibfield  {journal} {\bibinfo  {journal} {Phys.
  Rev. C}\ }\textbf {\bibinfo {volume} {98}},\ \bibinfo {pages} {014324}
  (\bibinfo {year} {2018})}\BibitemShut {NoStop}%
\bibitem [{\citenamefont {Simonis}\ \emph {et~al.}(2019)\citenamefont
  {Simonis}, \citenamefont {Bacca},\ and\ \citenamefont {Hagen}}]{simonis2019}%
  \BibitemOpen
  \bibfield  {author} {\bibinfo {author} {\bibfnamefont {J.}~\bibnamefont
  {Simonis}}, \bibinfo {author} {\bibfnamefont {S.}~\bibnamefont {Bacca}}, \
  and\ \bibinfo {author} {\bibfnamefont {G.}~\bibnamefont {Hagen}},\ }\bibfield
   {title} {\enquote {\bibinfo {title} {First principles electromagnetic
  responses in medium-mass nuclei - recent progress from coupled-cluster
  theory},}\ }\href {\doibase 10.1140/epja/i2019-12825-0} {\bibfield  {journal}
  {\bibinfo  {journal} {Eur. Phys. J. A}\ }\textbf {\bibinfo {volume} {55}},\
  \bibinfo {pages} {241} (\bibinfo {year} {2019})}\BibitemShut {NoStop}%
\bibitem [{\citenamefont {Sobczyk}\ \emph {et~al.}(2021)\citenamefont
  {Sobczyk}, \citenamefont {Acharya}, \citenamefont {Bacca},\ and\
  \citenamefont {Hagen}}]{Sobczyk:2021dwm}%
  \BibitemOpen
  \bibfield  {author} {\bibinfo {author} {\bibfnamefont {J.~E.}\ \bibnamefont
  {Sobczyk}}, \bibinfo {author} {\bibfnamefont {B.}~\bibnamefont {Acharya}},
  \bibinfo {author} {\bibfnamefont {S.}~\bibnamefont {Bacca}}, \ and\ \bibinfo
  {author} {\bibfnamefont {G.}~\bibnamefont {Hagen}},\ }\bibfield  {title}
  {\enquote {\bibinfo {title} {{Ab initio computation of the longitudinal
  response function in $^{40}$Ca}},}\ }\href {\doibase
  10.1103/PhysRevLett.127.072501} {\bibfield  {journal} {\bibinfo  {journal}
  {Phys. Rev. Lett.}\ }\textbf {\bibinfo {volume} {127}},\ \bibinfo {pages}
  {072501} (\bibinfo {year} {2021})},\ \Eprint
  {http://arxiv.org/abs/2103.06786} {arXiv:2103.06786 [nucl-th]} \BibitemShut
  {NoStop}%
\bibitem [{\citenamefont {Sobczyk}\ \emph
  {et~al.}(2024{\natexlab{a}})\citenamefont {Sobczyk}, \citenamefont {Acharya},
  \citenamefont {Bacca},\ and\ \citenamefont {Hagen}}]{Sobczyk:2023sxh}%
  \BibitemOpen
  \bibfield  {author} {\bibinfo {author} {\bibfnamefont {J.~E.}\ \bibnamefont
  {Sobczyk}}, \bibinfo {author} {\bibfnamefont {B.}~\bibnamefont {Acharya}},
  \bibinfo {author} {\bibfnamefont {S.}~\bibnamefont {Bacca}}, \ and\ \bibinfo
  {author} {\bibfnamefont {G.}~\bibnamefont {Hagen}},\ }\bibfield  {title}
  {\enquote {\bibinfo {title} {{Ca40 transverse response function from
  coupled-cluster theory}},}\ }\href {\doibase 10.1103/PhysRevC.109.025502}
  {\bibfield  {journal} {\bibinfo  {journal} {Phys. Rev. C}\ }\textbf {\bibinfo
  {volume} {109}},\ \bibinfo {pages} {025502} (\bibinfo {year}
  {2024}{\natexlab{a}})},\ \Eprint {http://arxiv.org/abs/2310.03109}
  {arXiv:2310.03109 [nucl-th]} \BibitemShut {NoStop}%
\bibitem [{\citenamefont {Jiang}\ \emph {et~al.}(2020)\citenamefont {Jiang},
  \citenamefont {Ekstr\"om}, \citenamefont {Forss\'en}, \citenamefont {Hagen},
  \citenamefont {Jansen},\ and\ \citenamefont {Papenbrock}}]{jiang2020}%
  \BibitemOpen
  \bibfield  {author} {\bibinfo {author} {\bibfnamefont {W.~G.}\ \bibnamefont
  {Jiang}}, \bibinfo {author} {\bibfnamefont {A.}~\bibnamefont {Ekstr\"om}},
  \bibinfo {author} {\bibfnamefont {C.}~\bibnamefont {Forss\'en}}, \bibinfo
  {author} {\bibfnamefont {G.}~\bibnamefont {Hagen}}, \bibinfo {author}
  {\bibfnamefont {G.~R.}\ \bibnamefont {Jansen}}, \ and\ \bibinfo {author}
  {\bibfnamefont {T.}~\bibnamefont {Papenbrock}},\ }\bibfield  {title}
  {\enquote {\bibinfo {title} {Accurate bulk properties of nuclei from $a=2$ to
  $\ensuremath{\infty}$ from potentials with $\mathrm{\ensuremath{\Delta}}$
  isobars},}\ }\href {\doibase 10.1103/PhysRevC.102.054301} {\bibfield
  {journal} {\bibinfo  {journal} {Phys. Rev. C}\ }\textbf {\bibinfo {volume}
  {102}},\ \bibinfo {pages} {054301} (\bibinfo {year} {2020})}\BibitemShut
  {NoStop}%
\bibitem [{\citenamefont {O'Connell}\ \emph {et~al.}(1987)\citenamefont
  {O'Connell} \emph {et~al.}}]{OConnell:1987ag}%
  \BibitemOpen
  \bibfield  {author} {\bibinfo {author} {\bibfnamefont {J.~S.}\ \bibnamefont
  {O'Connell}} \emph {et~al.},\ }\bibfield  {title} {\enquote {\bibinfo {title}
  {Electromagnetic excitation of the delta resonance in nuclei},}\ }\href
  {\doibase 10.1103/PhysRevC.35.1063} {\bibfield  {journal} {\bibinfo
  {journal} {Phys. Rev.}\ }\textbf {\bibinfo {volume} {C35}},\ \bibinfo {pages}
  {1063} (\bibinfo {year} {1987})}\BibitemShut {NoStop}%
\bibitem [{\citenamefont {Anghinolfi}\ \emph {et~al.}(1996)\citenamefont
  {Anghinolfi} \emph {et~al.}}]{Anghinolfi:1996vm}%
  \BibitemOpen
  \bibfield  {author} {\bibinfo {author} {\bibfnamefont {M.}~\bibnamefont
  {Anghinolfi}} \emph {et~al.},\ }\bibfield  {title} {\enquote {\bibinfo
  {title} {Quasi-elastic and inelastic inclusive electron scattering from an
  oxygen jet target},}\ }\href@noop {} {\bibfield  {journal} {\bibinfo
  {journal} {Nucl. Phys.}\ }\textbf {\bibinfo {volume} {A602}},\ \bibinfo
  {pages} {405--422} (\bibinfo {year} {1996})},\ \Eprint
  {http://arxiv.org/abs/nucl-th/9603001} {nucl-th/9603001} \BibitemShut
  {NoStop}%
\bibitem [{\citenamefont {Acharya}\ and\ \citenamefont
  {Bacca}(2022)}]{Acharya:2021lrv}%
  \BibitemOpen
  \bibfield  {author} {\bibinfo {author} {\bibfnamefont {Bijaya}\ \bibnamefont
  {Acharya}}\ and\ \bibinfo {author} {\bibfnamefont {Sonia}\ \bibnamefont
  {Bacca}},\ }\bibfield  {title} {\enquote {\bibinfo {title} {{Gaussian process
  error modeling for chiral effective-field-theory calculations of
  np\ensuremath{\leftrightarrow}d\ensuremath{\gamma} at low energies}},}\
  }\href {\doibase 10.1016/j.physletb.2022.137011} {\bibfield  {journal}
  {\bibinfo  {journal} {Phys. Lett. B}\ }\textbf {\bibinfo {volume} {827}},\
  \bibinfo {pages} {137011} (\bibinfo {year} {2022})},\ \Eprint
  {http://arxiv.org/abs/2109.13972} {arXiv:2109.13972 [nucl-th]} \BibitemShut
  {NoStop}%
\bibitem [{\citenamefont {Melendez}\ \emph {et~al.}(2019)\citenamefont
  {Melendez}, \citenamefont {Furnstahl}, \citenamefont {Phillips},
  \citenamefont {Pratola},\ and\ \citenamefont
  {Wesolowski}}]{Melendez:2019izc}%
  \BibitemOpen
  \bibfield  {author} {\bibinfo {author} {\bibfnamefont {J.~A.}\ \bibnamefont
  {Melendez}}, \bibinfo {author} {\bibfnamefont {R.~J.}\ \bibnamefont
  {Furnstahl}}, \bibinfo {author} {\bibfnamefont {D.~R.}\ \bibnamefont
  {Phillips}}, \bibinfo {author} {\bibfnamefont {M.~T.}\ \bibnamefont
  {Pratola}}, \ and\ \bibinfo {author} {\bibfnamefont {S.}~\bibnamefont
  {Wesolowski}},\ }\bibfield  {title} {\enquote {\bibinfo {title} {{Quantifying
  Correlated Truncation Errors in Effective Field Theory}},}\ }\href {\doibase
  10.1103/PhysRevC.100.044001} {\bibfield  {journal} {\bibinfo  {journal}
  {Phys. Rev. C}\ }\textbf {\bibinfo {volume} {100}},\ \bibinfo {pages}
  {044001} (\bibinfo {year} {2019})},\ \Eprint
  {http://arxiv.org/abs/1904.10581} {arXiv:1904.10581 [nucl-th]} \BibitemShut
  {NoStop}%
\bibitem [{\citenamefont {Ekstr\"om}\ \emph {et~al.}(2015)\citenamefont
  {Ekstr\"om}, \citenamefont {Jansen}, \citenamefont {Wendt}, \citenamefont
  {Hagen}, \citenamefont {Papenbrock}, \citenamefont {Carlsson}, \citenamefont
  {Forss\'en}, \citenamefont {Hjorth-Jensen}, \citenamefont {Navr\'atil},\ and\
  \citenamefont {Nazarewicz}}]{Ekstrom:2015rta}%
  \BibitemOpen
  \bibfield  {author} {\bibinfo {author} {\bibfnamefont {A.}~\bibnamefont
  {Ekstr\"om}}, \bibinfo {author} {\bibfnamefont {G.~R.}\ \bibnamefont
  {Jansen}}, \bibinfo {author} {\bibfnamefont {K.~A.}\ \bibnamefont {Wendt}},
  \bibinfo {author} {\bibfnamefont {G.}~\bibnamefont {Hagen}}, \bibinfo
  {author} {\bibfnamefont {T.}~\bibnamefont {Papenbrock}}, \bibinfo {author}
  {\bibfnamefont {B.~D.}\ \bibnamefont {Carlsson}}, \bibinfo {author}
  {\bibfnamefont {C.}~\bibnamefont {Forss\'en}}, \bibinfo {author}
  {\bibfnamefont {M.}~\bibnamefont {Hjorth-Jensen}}, \bibinfo {author}
  {\bibfnamefont {P.}~\bibnamefont {Navr\'atil}}, \ and\ \bibinfo {author}
  {\bibfnamefont {W.}~\bibnamefont {Nazarewicz}},\ }\bibfield  {title}
  {\enquote {\bibinfo {title} {{Accurate nuclear radii and binding energies
  from a chiral interaction}},}\ }\href {\doibase 10.1103/PhysRevC.91.051301}
  {\bibfield  {journal} {\bibinfo  {journal} {Phys. Rev. C}\ }\textbf {\bibinfo
  {volume} {91}},\ \bibinfo {pages} {051301} (\bibinfo {year} {2015})},\
  \Eprint {http://arxiv.org/abs/1502.04682} {arXiv:1502.04682 [nucl-th]}
  \BibitemShut {NoStop}%
\bibitem [{\citenamefont {Acharya}(2024)}]{Acharya:2024col}%
  \BibitemOpen
  \bibfield  {author} {\bibinfo {author} {\bibfnamefont {Bijaya}\ \bibnamefont
  {Acharya}},\ }\bibfield  {title} {\enquote {\bibinfo {title} {{Uncertainty
  Estimation and Anomaly Detection in Chiral Effective Field Theory Studies of
  Key Nuclear Electroweak Processes}},}\ }\href {\doibase
  10.1007/s00601-024-01929-x} {\bibfield  {journal} {\bibinfo  {journal} {Few
  Body Syst.}\ }\textbf {\bibinfo {volume} {65}},\ \bibinfo {pages} {65}
  (\bibinfo {year} {2024})},\ \Eprint {http://arxiv.org/abs/2404.11522}
  {arXiv:2404.11522 [nucl-th]} \BibitemShut {NoStop}%
\bibitem [{\citenamefont {Phillips}(2016)}]{Phillips:2016mov}%
  \BibitemOpen
  \bibfield  {author} {\bibinfo {author} {\bibfnamefont {Daniel~R.}\
  \bibnamefont {Phillips}},\ }\bibfield  {title} {\enquote {\bibinfo {title}
  {{Electromagnetic Structure of Two- and Three-Nucleon Systems: An Effective
  Field Theory Description}},}\ }\href {\doibase
  10.1146/annurev-nucl-102014-022321} {\bibfield  {journal} {\bibinfo
  {journal} {Ann. Rev. Nucl. Part. Sci.}\ }\textbf {\bibinfo {volume} {66}},\
  \bibinfo {pages} {421--447} (\bibinfo {year} {2016})}\BibitemShut {NoStop}%
\bibitem [{\citenamefont {Sobczyk}\ and\ \citenamefont
  {Bacca}(2024)}]{Sobczyk:2023mey}%
  \BibitemOpen
  \bibfield  {author} {\bibinfo {author} {\bibfnamefont {Joanna~E.}\
  \bibnamefont {Sobczyk}}\ and\ \bibinfo {author} {\bibfnamefont {Sonia}\
  \bibnamefont {Bacca}},\ }\bibfield  {title} {\enquote {\bibinfo {title} {{O16
  spectral function from coupled-cluster theory: Applications to lepton-nucleus
  scattering}},}\ }\href {\doibase 10.1103/PhysRevC.109.044314} {\bibfield
  {journal} {\bibinfo  {journal} {Phys. Rev. C}\ }\textbf {\bibinfo {volume}
  {109}},\ \bibinfo {pages} {044314} (\bibinfo {year} {2024})},\ \Eprint
  {http://arxiv.org/abs/2309.00355} {arXiv:2309.00355 [nucl-th]} \BibitemShut
  {NoStop}%
\bibitem [{\citenamefont {Wang}\ \emph {et~al.}(2012)\citenamefont {Wang},
  \citenamefont {Audi}, \citenamefont {Wapstra}, \citenamefont {Kondev},
  \citenamefont {MacCormick}, \citenamefont {Xu},\ and\ \citenamefont
  {Pfeiffer}}]{Wang:2012eof}%
  \BibitemOpen
  \bibfield  {author} {\bibinfo {author} {\bibfnamefont {M.}~\bibnamefont
  {Wang}}, \bibinfo {author} {\bibfnamefont {G.}~\bibnamefont {Audi}}, \bibinfo
  {author} {\bibfnamefont {A.~H.}\ \bibnamefont {Wapstra}}, \bibinfo {author}
  {\bibfnamefont {F.~G.}\ \bibnamefont {Kondev}}, \bibinfo {author}
  {\bibfnamefont {M.}~\bibnamefont {MacCormick}}, \bibinfo {author}
  {\bibfnamefont {X.}~\bibnamefont {Xu}}, \ and\ \bibinfo {author}
  {\bibfnamefont {B.}~\bibnamefont {Pfeiffer}},\ }\bibfield  {title} {\enquote
  {\bibinfo {title} {{The Ame2012 atomic mass evaluation}},}\ }\href {\doibase
  10.1088/1674-1137/36/12/003} {\bibfield  {journal} {\bibinfo  {journal}
  {Chin. Phys. C}\ }\textbf {\bibinfo {volume} {36}},\ \bibinfo {pages}
  {1603--2014} (\bibinfo {year} {2012})}\BibitemShut {NoStop}%
\bibitem [{\citenamefont {Epelbaum}\ \emph {et~al.}(2014)\citenamefont
  {Epelbaum}, \citenamefont {Krebs}, \citenamefont {L\"ahde}, \citenamefont
  {Lee}, \citenamefont {Mei\ss{}ner},\ and\ \citenamefont
  {Rupak}}]{epelbaum2014}%
  \BibitemOpen
  \bibfield  {author} {\bibinfo {author} {\bibfnamefont {Evgeny}\ \bibnamefont
  {Epelbaum}}, \bibinfo {author} {\bibfnamefont {Hermann}\ \bibnamefont
  {Krebs}}, \bibinfo {author} {\bibfnamefont {Timo~A.}\ \bibnamefont
  {L\"ahde}}, \bibinfo {author} {\bibfnamefont {Dean}\ \bibnamefont {Lee}},
  \bibinfo {author} {\bibfnamefont {Ulf-G.}\ \bibnamefont {Mei\ss{}ner}}, \
  and\ \bibinfo {author} {\bibfnamefont {Gautam}\ \bibnamefont {Rupak}},\
  }\bibfield  {title} {\enquote {\bibinfo {title} {Ab initio calculation of the
  spectrum and structure of $^{16}\mathrm{O}$},}\ }\href {\doibase
  10.1103/PhysRevLett.112.102501} {\bibfield  {journal} {\bibinfo  {journal}
  {Phys. Rev. Lett.}\ }\textbf {\bibinfo {volume} {112}},\ \bibinfo {pages}
  {102501} (\bibinfo {year} {2014})}\BibitemShut {NoStop}%
\bibitem [{\citenamefont {Gour}\ \emph {et~al.}(2006)\citenamefont {Gour},
  \citenamefont {Piecuch}, \citenamefont {Hjorth-Jensen}, \citenamefont
  {W\l{}och},\ and\ \citenamefont {Dean}}]{gour2006}%
  \BibitemOpen
  \bibfield  {author} {\bibinfo {author} {\bibfnamefont {J.~R.}\ \bibnamefont
  {Gour}}, \bibinfo {author} {\bibfnamefont {P.}~\bibnamefont {Piecuch}},
  \bibinfo {author} {\bibfnamefont {M.}~\bibnamefont {Hjorth-Jensen}}, \bibinfo
  {author} {\bibfnamefont {M.}~\bibnamefont {W\l{}och}}, \ and\ \bibinfo
  {author} {\bibfnamefont {D.~J.}\ \bibnamefont {Dean}},\ }\bibfield  {title}
  {\enquote {\bibinfo {title} {Coupled-cluster calculations for valence systems
  around $^{16}\mathrm{O}$},}\ }\href {\doibase 10.1103/PhysRevC.74.024310}
  {\bibfield  {journal} {\bibinfo  {journal} {Phys. Rev. C}\ }\textbf {\bibinfo
  {volume} {74}},\ \bibinfo {pages} {024310} (\bibinfo {year}
  {2006})}\BibitemShut {NoStop}%
\bibitem [{\citenamefont {Sobczyk}\ \emph
  {et~al.}(2024{\natexlab{b}})\citenamefont {Sobczyk}, \citenamefont {Rocco},\
  and\ \citenamefont {Lovato}}]{Sobczyk:ANN}%
  \BibitemOpen
  \bibfield  {author} {\bibinfo {author} {\bibfnamefont {Joanna~E.}\
  \bibnamefont {Sobczyk}}, \bibinfo {author} {\bibfnamefont {Noemi}\
  \bibnamefont {Rocco}}, \ and\ \bibinfo {author} {\bibfnamefont {Alessandro}\
  \bibnamefont {Lovato}},\ }\bibfield  {title} {\enquote {\bibinfo {title}
  {{Modeling inclusive electron-nucleus scattering with Bayesian artificial
  neural networks}},}\ }\href@noop {} {\  (\bibinfo {year}
  {2024}{\natexlab{b}})},\ \Eprint {http://arxiv.org/abs/2406.06292}
  {arXiv:2406.06292 [nucl-th]} \BibitemShut {NoStop}%
\end{thebibliography}%

\end{document}